\documentclass{emulateapj}
\usepackage{amssymb}
\usepackage{apjfonts}
\usepackage{graphicx}
\usepackage{natbib}
\usepackage{rotating, multirow}
\usepackage{lscape}

\def\CIII{C\,{\sevenrm\,III]}}
\def\CIV{C\,{\sevenrm\,IV}}

\def\MgII{Mg\,{\sevenrm\,II}}

\def\hbeta{H{$\beta$}}
\def\OIII{[O\,{\sevenrm\,III}]}

\newcommand{\OIIIb}{[O{\sevenrm\, III}]\,$\lambda$5007}
\newcommand{\OIIIab}{[O{\sevenrm\, III}]\,$\lambda\lambda$4959,5007}

\def\halpha{H{$\alpha$}}
\def\NII{[N\,{\sevenrm\,II}]}

\newcommand{\NIIb}{[N{\sevenrm\, II}]\,$\lambda$6585}
\newcommand{\NIIab}{[N{\sevenrm\, II}]\,$\lambda\lambda$6549,6585}

\def\SII{[S\,{\sevenrm\,II}]}

\newcommand{\SIIab}{[S{\sevenrm\, II}]\,$\lambda\lambda$6718,6732}

\def\FeII{Fe\,{\sevenrm\, II}}

\def\zem{$z_{\rm em}$}
\def\kms{$\rm km\,s^{-1}$}

 \font\sevenrm=cmr7 scaled 1000

\begin{document}
\title{A catalog of quasar properties from the Baryon Oscillation Spectroscopic Survey}
\shorttitle{A catalog of quasar properties from the BOSS}
\shortauthors{Chen et al.}

\author{Zhi-Fu Chen\altaffilmark{1}, Da-Sheng Pan\altaffilmark{2}, Ting-Ting Pang\altaffilmark{1}, Yong Huang\altaffilmark{1}}

\altaffiltext{1}{Department of Physics and Telecommunication Engineering, Baise University, Baise 533000, China; zhichenfu@126.com}
\altaffiltext{2}{Department of Information Technology, Guangxi Financial Vocational College, Nanning 530007, China}

\begin{abstract}
Using the quasars with $z_{\rm em} < 0.9$ from the Baryon Oscillation Spectroscopic Survey, we measure the spectral characteristics, including continuum and emission lines, around the \hbeta\ and \halpha\ spectral regions, which are lacking in Quasar Data Release 12 (DR12Q). We estimate the virial black hole mass from broad \halpha\ and/or \hbeta, and infer quasar redshifts from \OIIIb\ emission lines. All the measurements and derived quantities are publicly available. The comparison between \OIIIb\ based redshifts and the visual inspection redshifts included in DR12Q indicates that the visual inspection redshifts are robust. We find that the FWHMs of the broad \halpha\ are consistent with those of the broad \hbeta, while both the equivalent widths and line luminosities of the broad \halpha\ are obviously larger than the corresponding quantities of the broad \hbeta. We also find that there is an obviously systematic offset between the \hbeta\ and \halpha\ based mass if they are inferred from the empirical relationships in the literature. Using our large quasar sample, we have improved the \hbeta\ and \halpha\ based mass estimators by minimizing the difference between the \hbeta\- and \halpha\-based mass. For the black hole mass estimator (Equation (1)), we find that the coefficients $\rm (a,b)=(7.00,0.50)$ for the \halpha\ and $\rm (a,b)=(6.96,0.50)$ for the \hbeta\ are the best choices.
\end{abstract}

\section{Introduction}
It is widely accepted that supermassive black holes (SMBHs) reside at the central region of all massive galaxies, mostly in the form of active galactic nuclei \cite[AGNs; e.g.][]{2013ARA&A..51..511K}. The mass of the SMBH ($M_{\rm BH}$) can reach up to $10^{10}~M_{\odot}$ \cite[e.g.,][]{2015Natur.518..512W}. A foundational and important problem is how the black hole evolves with cosmic time and what role the black hole plays in the properties and evolution of the quasar host galaxy and surrounding environment. It is found that the properties of SMHBs, such as accretion rate density and mass, are tightly connected to the global properties of the stellar within their host galaxies \cite[e.g.,][]{2009MNRAS.397.1705G,2013ARA&A..51..511K,2014ARA&A..52..415M}. However, the underlying physics is still widely debated, particularly why, when and how the SMBHs and their host galaxies regulate one another. A crucial issue in these problems is measuring the emission characteristics and $M_{\rm BH}$ of AGNs/quasars.

Assuming a virial broad emission line region ($\rm BLR$), using the measurement of broad emission line width ($\upsilon$) and the distance $R$ between the central region and the $\rm BLR$, which can be revealed by reverberation mapping \cite[e.g.;][]{1982ApJ...255..419B,1993PASP..105..247P}, we can determine the black hole mass $M_{\rm BH}=R\upsilon^2/G$, where $G$ is the gravitational constant. With the help of the empirical relationship between $R$ and luminosity ($L$) of the central region \cite[e.g.;][]{2000ApJ...533..631K}, one can easily obtain $M_{\rm BH}$ for a large number of objects with the $L$ and $\upsilon$ which can be directly yielded from a single-epoch spectrum.

The Sloan Digital Sky survey \cite[SDSS;][]{2000AJ....120.1579Y} is a great and ambitious project in astronomy. During the third stage of the SDSS, the Baryon Oscillation Spectroscopic Survey \cite[BOSS;][]{2013AJ....145...10D} has obtained $\rm 297~301$ unique quasar spectra \cite[][]{2017A&A...597A..79P}, offering a great opportunity to infer the properties of quasars. Broad \CIV, \MgII, \hbeta\ and \halpha\ emission lines are important tools for describing the dynamic mechanisms within the broad emission line region of quasars, whose characteristics are also crucial to infer the properties of SMBHs. \cite{2017A&A...597A..79P} have measured the full width at half maximum (FWHM) for the broad \CIV\ and \MgII\ emission lines, but lack the continuum luminosities and emission measurements in \hbeta\ spectral region (including \hbeta\ and \OIIIab) and \halpha\ spectral region (including \halpha, \NIIab\ and \SIIab). The continuum luminosity is crucial to infer the $M_{\rm BH}$ of the SMBHs. Using the FWHMs of \CIV\ and \MgII\ from \cite{2017A&A...597A..79P} and the continuum luminosities that are derived from the broadband urgiz magnitudes of the SDSS \cite[][]{1996AJ....111.1748F}, \cite{2017ApJS..228....9K} estimated the black hole masses for most of the quasars included in \cite{2017A&A...597A..79P}. While, both \cite{2017ApJS..228....9K} and \cite{2017A&A...597A..79P} lack the measurements in \hbeta\ and \halpha\ spectral regions, which are very important to constrain the properties of quasars, especially for the quasars with low redshifts. In this paper, we make up these deficiencies by analyzing the BOSS spectral data in \hbeta\ and \halpha\ spectral regions.

In Section \ref{sect:datasample}, we describe the quasar sample, spectral measurements, and catalogs. We present the collective properties of the measurements and discussions in Section \ref{sect:discussion}. The  summary is presented in Section \ref{sect:summary}. In this paper, we adopt the $\rm \Lambda CDM$ cosmology with $\rm \Omega_M=0.3$, $\rm \Omega_\Lambda=0.7$, and $\rm H_0=70~km~s^{-1}~Mpc^{-1}$.

\section{The quasar sample and spectral measurements}
\label{sect:datasample}
The SDSS uses a dedicated wide-field 2.5 m telescope \cite[][]{2006AJ....131.2332G} with drift-scan camera \cite[][]{1998AJ....116.3040G}, located at Apache Point Observatory, New Mexico to image the universe in five broad bands \cite[ugriz;][]{1996AJ....111.1748F}. The SDSS obtained the first light in 1998 May, and operated regular survey in 2000. The BOSS is the main dark time legacy survey of the third phase of the SDSS \cite[SDSS-III;][]{2011AJ....142...72E} from 2008 July to 2014 June, who gathered quasar spectra data \cite[][]{2012ApJS..199....3R} under the guideline of the SDSS-DR8 imaging data \cite[][]{2011ApJS..193...29A}. The BOSS produces spectra in a wavelength range of 3600 \AA\ $< \lambda<$ 10400 \AA\ at a resolution of $R=1300\sim2500$ \cite[][]{2015ApJS..219...12A}.

The Data Release 12 Quasar Catalog of the SDSS \cite[DR12Q;][]{2017A&A...597A..79P} includes $\rm 297~301$ unique quasars, which are detected in a sky area over 9376 $\rm deg^2$. These quasars are visually and spectroscopically confirmed, brighter than $M_{\rm  i}(z=2) = -20.5$, and have at least one emission line with an FWHM $>500$ \kms\ or interesting/complex absorption features. Most of quasars (184101/297301) are located at $z>2.15$. \cite{2017A&A...597A..79P} have measured the emission characteristics of quasars for \CIV, \CIII\ and/or \MgII\ lines, while lack the measurements of \hbeta, \OIIIab, \halpha, \NIIab\ and \SIIab\ emission lines, which are important to limit the properties of quasar supermassive black hole, accretion disc, broad emission line region, narrow emission line region and host galaxy, especially for the quasars with low redshifts whose UV emission lines are not available in the SDSS spectra. This paper will measure the characteristics of \hbeta, \OIII, \halpha, \NII\ and \SII\ emission lines for the quasars included in DR12Q.

The spectral fitting methods used in this paper are similar to those utilized in \cite{2011ApJS..194...45S} and \cite{2009MNRAS.397.1713C}. We correct the Galactic extinction in the BOSS spectra based on the reddening measurements of \cite{2011ApJ...737..103S} and the Milky Way extinction curve from \cite{1989ApJ...345..245C}. We only focus on the quasar emissions in the \hbeta\ spectral region (including \hbeta\ and \OIII\ emission lines) and the \halpha\ spectral region (including \halpha, \NII\ and \SII\ emission lines). For each quasar, we fit local power-law continuum ($f_{\lambda}=A\lambda^{\alpha}$) plus iron template \cite[][]{2001ApJS..134....1V,2004A&A...417..515V} in \hbeta\ and \halpha\ spectral regions, respectively. These fits lack the contaminations of broad emission lines. The spectra subtracted by continuum+iron fits are invoked to measure emission characteristics with multi-Gaussian functions. In the following, we describe in detail the fitting programs in the \hbeta\ and \halpha\ spectral regions, respectively. The codes used to fit spectra are free to readers who contact us.

\subsection{H$\alpha$ spectral region}
\label{sect:halpha}
We only consider the quasars with $z_{\rm} \le0.4857$ and $\rm S/N >2~pixel^{-1}$, where the S/N is the median signal-to-noise ratio in the wavelength range from 6300 \AA\ to 6900 \AA. The spectral windows for the continuum+iron fits are [6000,6300] \AA\ and [6750,7000] \AA, and the fits are performed with decreasing $\chi^2$. The emission-line-fitting program is for spectral data at wavelength range from 6400 \AA\ to 6800 \AA, which contain broad and narrow \halpha\ components, a narrow \NIIab\ doublet, and a narrow \SIIab\ doublet. Each narrow component is described by a single Gaussian function and is constrained with a line width FWHM $<1200$ \kms. For all narrow components, we tie together their velocity offsets from the visual inspection redshift of the quasar and force their line widths to be the same. The flux ratio of the \NIIab\ doublet is fixed to be $\rm F(6585)/F(6549)=2.96$. The broad \halpha\ component is described with two different methods: (1) a single Gaussian function with a line width FWHM $>1200$ \kms; and (2) three Gaussian functions and each with a line width FWHM $>1200$ \kms. All the emission line components are fitted simultaneously with decreasing $\chi^2$.

\subsection{$H\beta$ spectral region}
\label{sect:hbeta}
We only consider the quasars with $z_{\rm} \le0.8909$ and $ S/N >2~pixel^{-1}$, where the S/N is the median signal-to-noise ratio in the wavelength range from 4700 \AA\ to 5100 \AA. The continuum+iron fitting windows are [4400,4800] \AA\ and [5100,5550] \AA. The emission-line-fitting program is performed on spectral data at wavelength range from 4700 \AA\ to 5100 \AA. The line-fitting process considers simultaneously broad and narrow \hbeta\ components, and the \OIIIab\ doublet. The narrow \hbeta\ component is described with a single Gaussian function with a line width FWHM $<1200$ \kms. It is similar to the case of broad \halpha\ emission that the broad \hbeta\ component is characterized in two different ways: either in a single Gaussian function or in three Gaussian functions, and each function with a line width FWHM $>1200$ \kms. The \OIII\ emissions often display complex line profiles. Outflow has become a fashionable and important component within a quasar system, which frequently drives \OIII\ emission features with asymmetric blue wings or even double-peaked profiles \cite[e.g.,][]{2008ApJ...680..926K,2010ApJ...708..427L}. Therefore, we invoke two Gaussian functions to characterize each component of a \OIIIab\ doublet, of which one is for the core and the other one is for the wing \cite[e.g.,][]{2011ApJS..194...45S,2014RAA....14..913P}. The flux ratio of the \OIIIab\ doublet is fixed to be $\rm F(5007)/F(4959)=3$. The velocity offset from the visual inspection redshift of the quasar and line width are constrained to be the same values for the narrow \hbeta\ and the core components of the \OIIIab\ doublet. For the wing components of the \OIIIab\ doublet, two lines of each doublet are imposed to be the same line width and velocity offset from the visual inspection redshift of the quasar. The wing components frequently show a wider line profiles relative to the core components. Therefore, we put a looser line width with FWHM $<2500$ \kms\ to limit the wing components of the \OIII\ emissions \cite[e.g.,][]{2014RAA....14..913P}.

\subsection{Discussions of the spectral fits}
\label{sect:Discussions_fits}
\begin{figure*}
\centering
\includegraphics[width=0.45\textwidth]{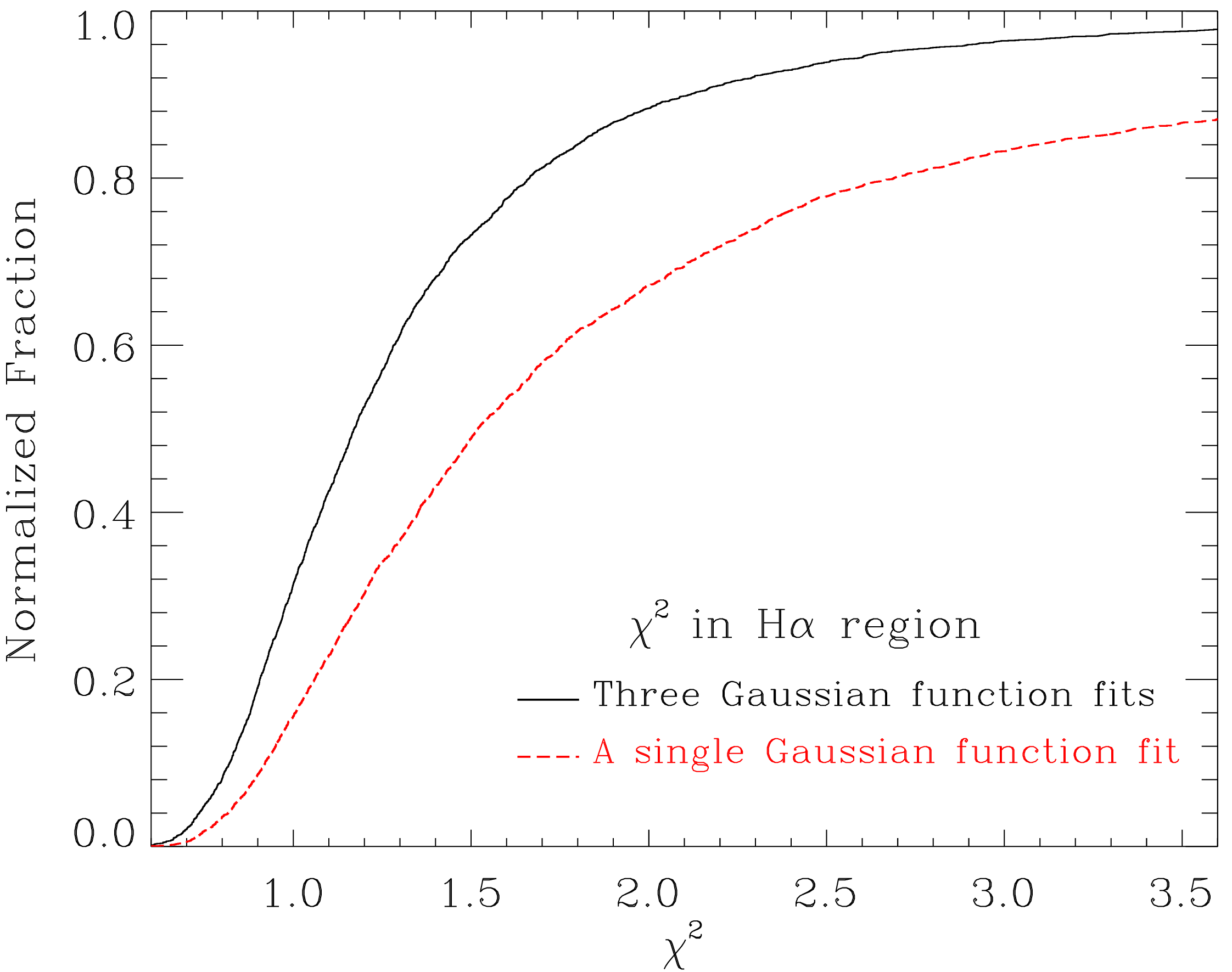}
\hspace{3ex}
\includegraphics[width=0.45\textwidth]{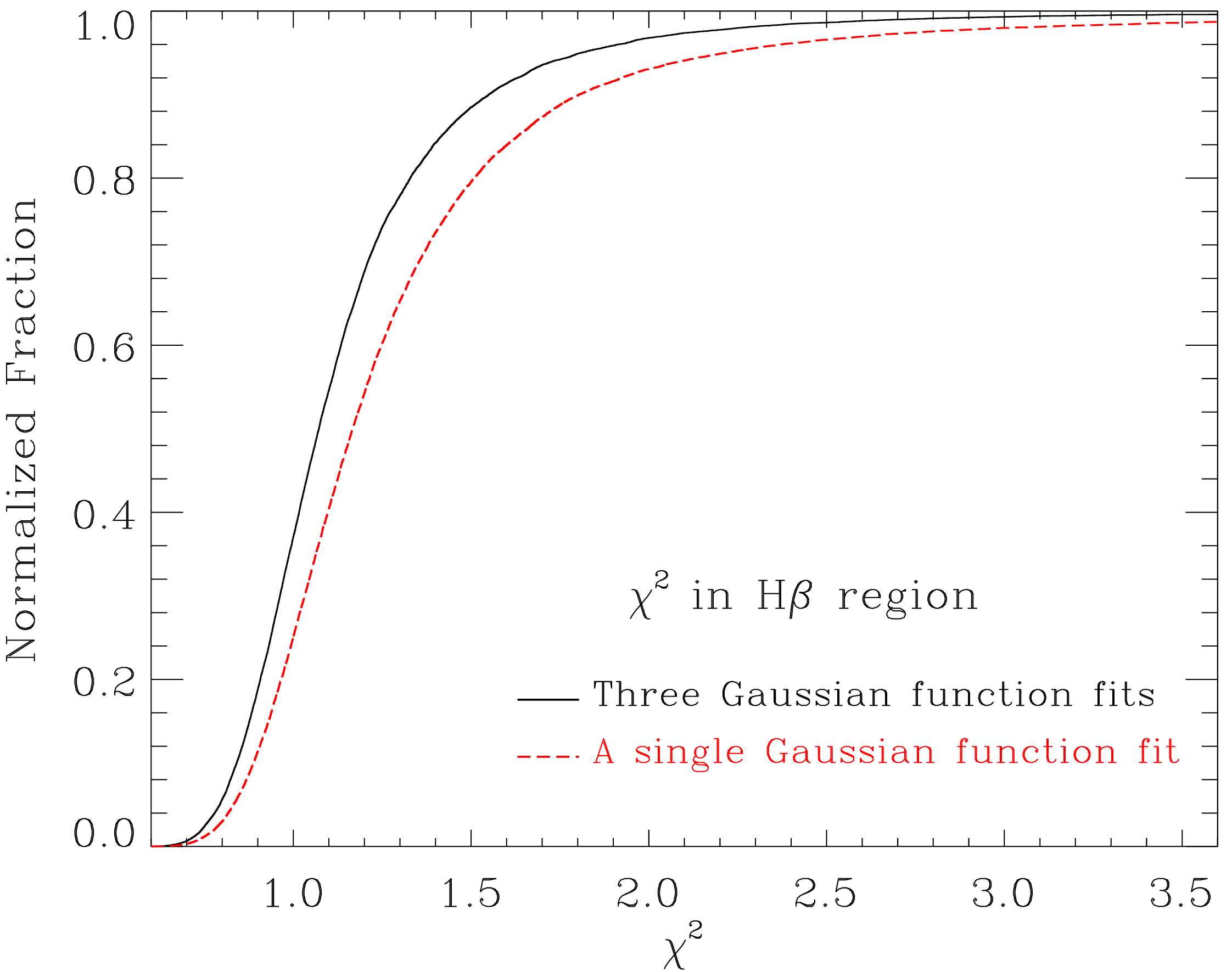}
\caption{Distributions of the fitting $\chi^2$. The y-axes have been normalized by the total number of quasars with line measurements. The black solid-lines mean the broad component with three Gaussian function fits, and the red dash lines represent the broad component with a single Gaussian function fits. The left panel shows the spectral fits in the \halpha\ region, and the right one is for the spectral fits in the \hbeta\ region. The three Gaussian function fits show a slightly smaller $\chi^2$ than a single Gaussian function fit does.}
\label{fig:chi_linefit}
\end{figure*}

The methods used to fit emission lines in \halpha\ and \hbeta\ spectral regions are consistent with those utilized in \cite{2011ApJS..194...45S}: (1) both a single Gaussian function and a combination of three Gaussian functions are invoked to model the broad \halpha\ or \hbeta\ emissions; (2) two pairs of Gaussian functions are used to model the \OIIIab\ emissions, one for the core component and the other for the wing one; and (3) a single Gaussian function for each of the narrow \hbeta, \NIIab\ and \SIIab\ lines. We note that a single Gaussian function can well describe a majority of broad \halpha\ or \hbeta\ components (see the $\chi^2$ distributions in Figure \ref{fig:chi_linefit}). And also, a combination of two Gaussian functions, one for the core component and the other one for the wing component, would well characterize a vast majority of the broad \halpha\ or \hbeta\ components. Nevertheless, a combination of three Gaussian functions is still invoked to model the broad components so that the methods used in this paper are consistent with those utilized in \cite{2011ApJS..194...45S}.

\begin{figure}
\centering
\includegraphics[width=0.45\textwidth]{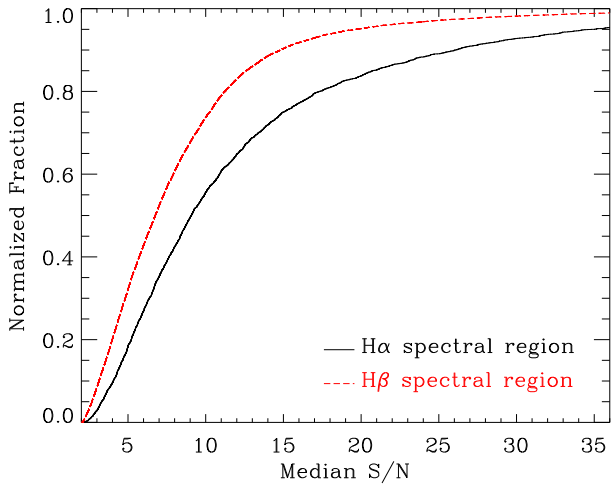}
\caption{Distributions of the median S/N per pixel around the line-fitting regions, which normalized by the total number of quasars with line measurements.}
\label{fig:Median_SNR}
\end{figure}

The fitting results are visually checked one-by-one. We are confident that the vast majority of the spectral fits ($>95\%$) are reliable. All the emission components in corresponding line-fitting regions are fitted simultaneously with decreasing $\chi^2$. We show the $\chi^2$ values in Figure \ref{fig:chi_linefit}. Relative to the fits describing the broad components with three Gaussian functions, the fits modeling the broad components with a single Gaussian function on average have a slightly larger $\chi^2$. Using three Gaussian functions to model the broad components, about 97\% of the fits in \halpha\ spectral regions and $>99\%$ of the fits in \hbeta\ spectral regions have a $\chi^2$ value of $<3$, which are similar to the fitting results of \cite{2011ApJS..194...45S}. The median S/N around line fitting regions for the quasars with line measurements is plotted in Figure \ref{fig:Median_SNR}, which shows that there are many quasar spectra with low median S/N. During the fitting processes, we fixed the flux ratios of the \OIIIab\ and \NIIab\ doublets to be constant values. However, we found that in many objects, the flux ratios deviated from the fixed values, which might be due to the more extreme physical conditions \cite[e.g.,][and references therein]{1985Msngr..39...15R,2000MNRAS.312..813S,2006MNRAS.366..480G}. The low S/N and the fixed flux ratios could reduce the goodness of the spectral fits for some quasars, and would be important reasons why some quasars modeling the broad components, even with three Gaussian functions, still exhibit large $\chi^2$ values.

\begin{figure*}
\centering
\includegraphics[width=0.45\textwidth]{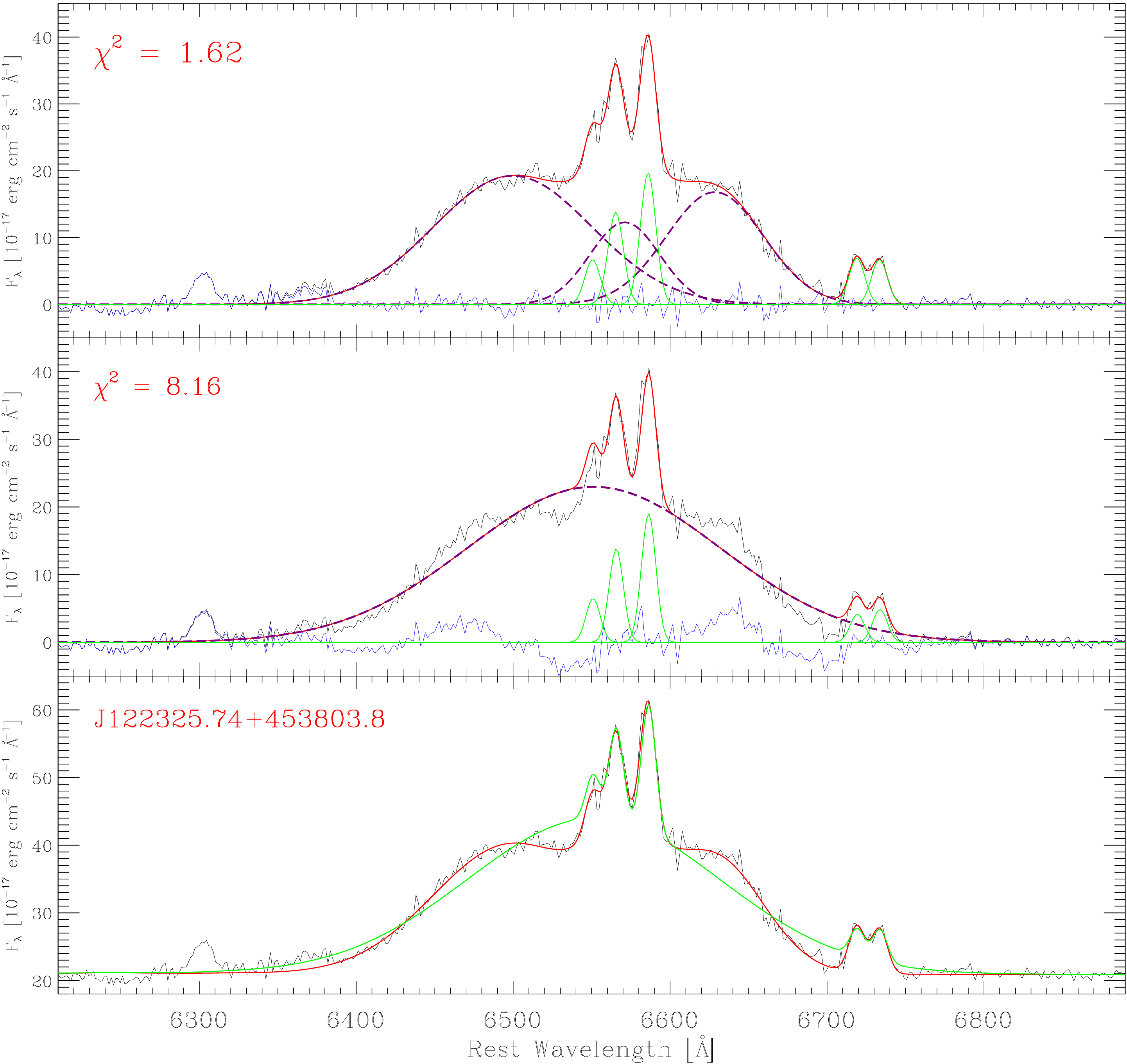}
\hspace{3ex}
\includegraphics[width=0.45\textwidth]{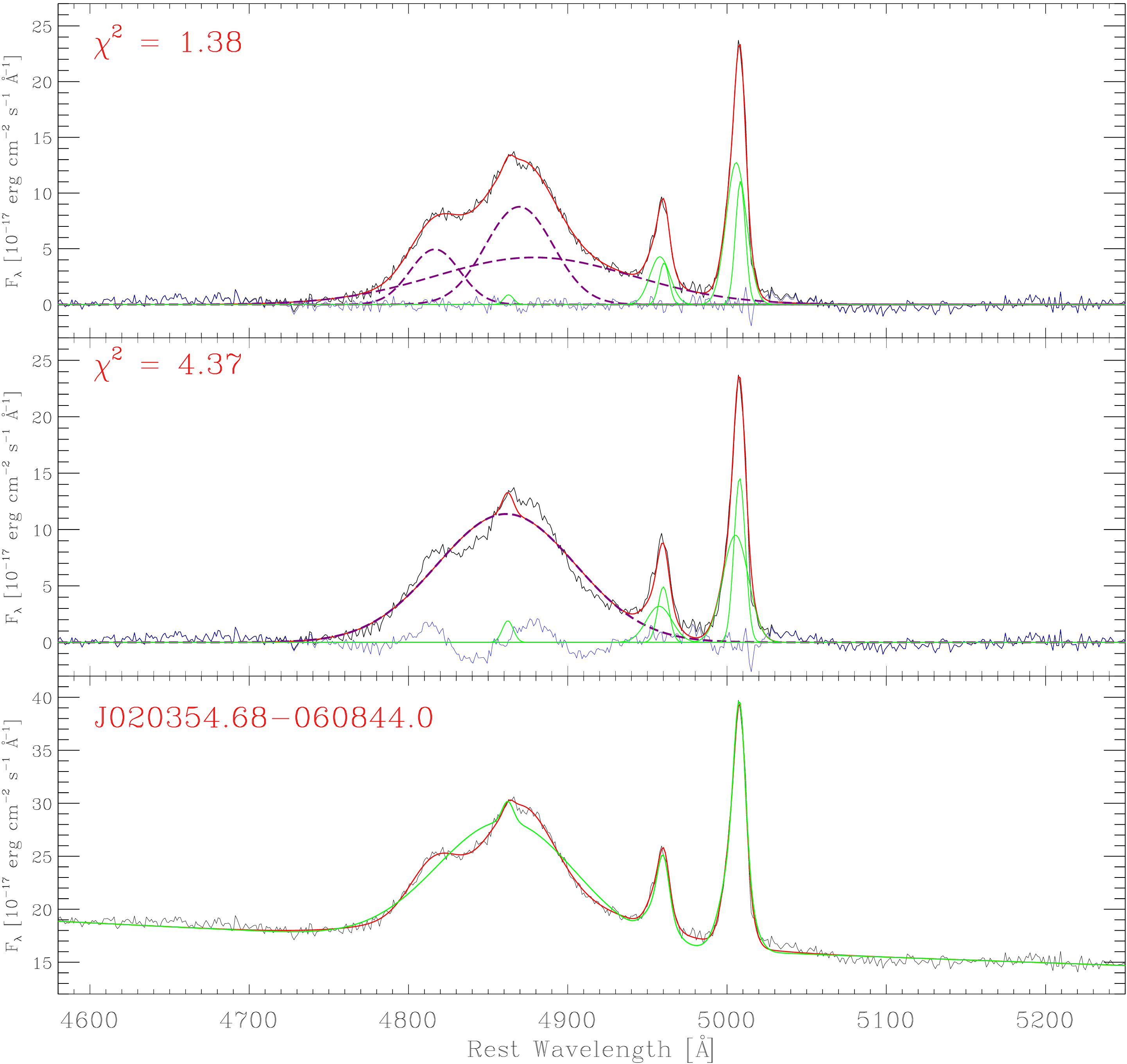}
\caption{Spectra of quasars J122325.74+453803.8 and J020354.68-060844.0. Bottom panels: spectra overplotted with fitting results. The green and red solid lines are for the fits using a single and three Gaussian functions to model broad components, respectively. Middle and upper panels: the black solid lines are the spectra subtracted by the sum of fitting the power-law continuum and \FeII\ emissions; the purple dash-lines represent the broad Gaussian components, the green solid lines represent the narrow Gaussian components; the red solid lines standard for the sum of all the fitting Gaussian components; and the blue solid lines are the residual fluxes which are the black solid lines subtracted by the red solid lines.}
\label{fig:samplespectra}
\end{figure*}

We find that with the broad component modeling from a single Gaussian function, about 17\% of the fits in \halpha\ spectral regions, and $\sim3\%$ of the fits in \hbeta\ spectral regions, have $\chi^2$ values of $>3$. These fractions are significantly larger than those of the fits modeling the broad components with three Gaussian functions. Outflow, inflow, and disk emitters could result in broad emission lines with asymmetric profiles, which, on average, cannot be as described with a single Gaussian function. As examples, Figure \ref{fig:samplespectra} compares the fits modeling broad emission components with a single and three Gaussian functions, respectively. It is clear that the broad components can be well described with a combination of three Gaussian functions, but not with a single Gaussian function.

\subsection{The black hole mass}
\label{sect:BHmass}
The reverberation mapping technique can infer the radius of the virialized broad emission line region (BLR) of the quasar from the time lag of variabilities between the continuum and broad emission lines \cite[e.g.;][]{1982ApJ...255..419B,1993PASP..105..247P}. Combining the $R_{\rm BRL}$ and the velocity dispersion of the BRL, one can derive the virial mass of the black hole ($M_{\rm BH}$) \cite[e.g.;][]{1999ApJ...526..579W,2000ApJ...533..631K}. Tight correlation between $R_{\rm BRL}$ and optical continuum luminosity largely reduces the difficulty of deriving $R_{\rm BRL}$ \cite[e.g.;][]{2000ApJ...533..631K}, which suggests that the optical continuum luminosity can be used as a proxy of the $R_{\rm BRL}$. The velocity dispersion of the BRL can be generally reflected by broad line width. One can directly measure the line width and continuum luminosity from a single-epoch spectrum and thereby derives the $M_{\rm BH}$. Using the spectral measurements, the $M_{\rm BH}$ is typically estimated via
\begin{equation}
\label{eq:BHmass_Hbeta}
 \rm Log (\frac{M_{BH}}{M_\odot}) = a + b\times Log(\frac{\lambda L_{\lambda}}{10^{44} erg~s^{-1}}) + 2\times Log(\frac{FWHM}{km~s^{-1}})
\end{equation}
where the $\lambda L_{\lambda}$ is the monochromatic luminosity of the continuum, and the calibrated coefficients (a,b) are empirical values \cite[e.g.;][]{2005ApJ...630..122G,2006ApJ...641..689V,2011ApJS..194...45S}, which are estimated typically by reverberation mapping of nearby AGNs.

In this paper, we utilize the \hbeta\ and/or \halpha\ broad emission lines and corresponding continuum luminosity at 5100 \AA\ ($L_{5100}$) to estimate the virial mass of the quasar central black hole, where the $L_{5100}$ is directly measured from the fitting power law. We use calibrated coefficients of $\rm (a,b)=(6.91,0.50)$ to infer the $M_{\rm BH}$ from \hbeta\ emission lines \cite[e.g.;][]{2006ApJ...641..689V,2011ApJS..194...45S}. For some low-redshift quasars with \halpha\ emission lines being available in the BOSS spectra, we also estimated the $M_{\rm BH}$ from \halpha\ emission lines. The total \halpha\ line luminosities are tightly related to the continuum luminosities at 5100 \AA, and the virial mass estimator based on the \halpha\ emissions can be expressed as \cite[][]{2005ApJ...630..122G}:
\begin{equation}
\label{eq:BHmass_Halpha}
 \rm Log (\frac{M_{BH}}{M_\odot}) = 0.12 + 0.55Log(\frac{L_{H\alpha}}{10^{42} erg~s^{-1}}) + 2.06Log(\frac{FWHM_{H\alpha}}{km~s^{-1}}).
\end{equation}
where $L_{\rm H\alpha}$ is the total luminosity of broad and narrow \halpha\ components. Section \ref{sect:discussion} will compares the $M_{\rm BH}$ obtained by \hbeta\ and \halpha\ emissions.

\subsection{The spectral catalog}
\label{sect:catalog}
We tabulate the measurements from spectral fits in the online catalogs of this paper. The descriptions of the catalogs are provided in Tables \ref{tab:halpha} and \ref{tab:hbeta}. Flux measurements have been corrected for galactic extinction based on the reddening measurements of \cite{2011ApJ...737..103S} and the Milky Way extinction curve from \cite{1989ApJ...345..245C}. We estimate the uncertainties of the equivalent widths and luminosities of detectable emission lines through integrating flux uncertainties within $\pm3\sigma$ line widths, where $\sigma=FWHM/2.3548$ is the velocity dispersion of the corresponding emission line. For broad emission components, we estimate the uncertainties with the FWHMs of the fits modeling the broad component with a single Gaussian function. One can approximately estimate the uncertainties with the FWHMs of the fits modeling the broad component with three Gaussian functions by comparing the FWHMs of the fits modeling the broad component with a single and three Gaussian functions.

\section{Collective properties and discussions}
\label{sect:discussion}
The spectral measurements of this paper are useful for investigating the statistical properties of quasars. In this section, we mainly present the collective properties of measurements.

\begin{figure}
\centering
\includegraphics[width=0.45\textwidth]{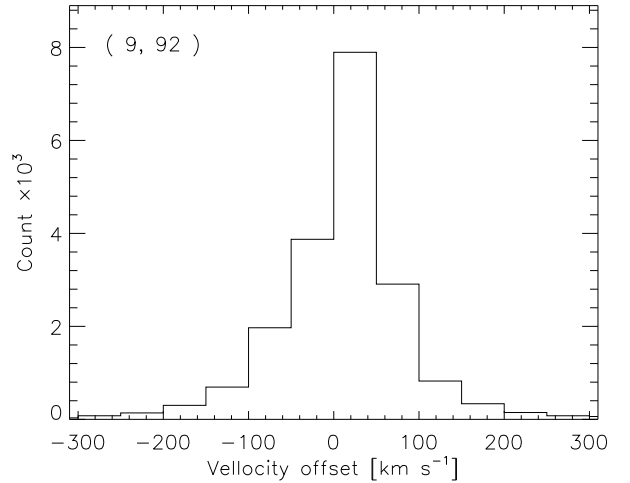}
\caption{Velocity offsets of the visual inspection redshifts ($\rm Z_{VI}$) included in DR12Q with respect to the \OIIIb\ based redshifts ($Z_{\rm \OIII}$) in this paper. The values in the top-left corner indicate the mean offset and the standard deviation.}
\label{fig:vrzz}
\end{figure}

Using a combination of multi eigenspectra that are inferred from a principal component analysis of the quasars with robust redshifts, the SDSS pipeline yields high accurate redshifts (pipeline redshifts) for most of quasars \cite[][]{2012AJ....144..144B}. In order to correct for the bad pipeline redshifts, \cite{2017A&A...597A..79P} visually checked all objects included in the DR12Q and modified the bad pipeline redshifts to \MgII\-based redshifts. Therefore, the visual inspection redshifts would be better values relative to the pipeline redshifts. Nevertheless, we note that the \MgII\-based redshifts would also show a bias with respect to the narrow \OIIIb\-based redshifts \cite[e.g.;][]{2016ApJ...817...55S}. Figure \ref{fig:vrzz} shows the velocity offsets of the visual inspection redshifts with respect to the \OIIIb\-based redshifts. We find a mean offset of 9 \kms\ and a standard deviation of 92 \kms\ for this velocity offset, which suggest that the visual inspection redshifts provided by \cite{2017A&A...597A..79P} are robust.

\begin{figure*}
\centering
\includegraphics[width=0.32\textwidth]{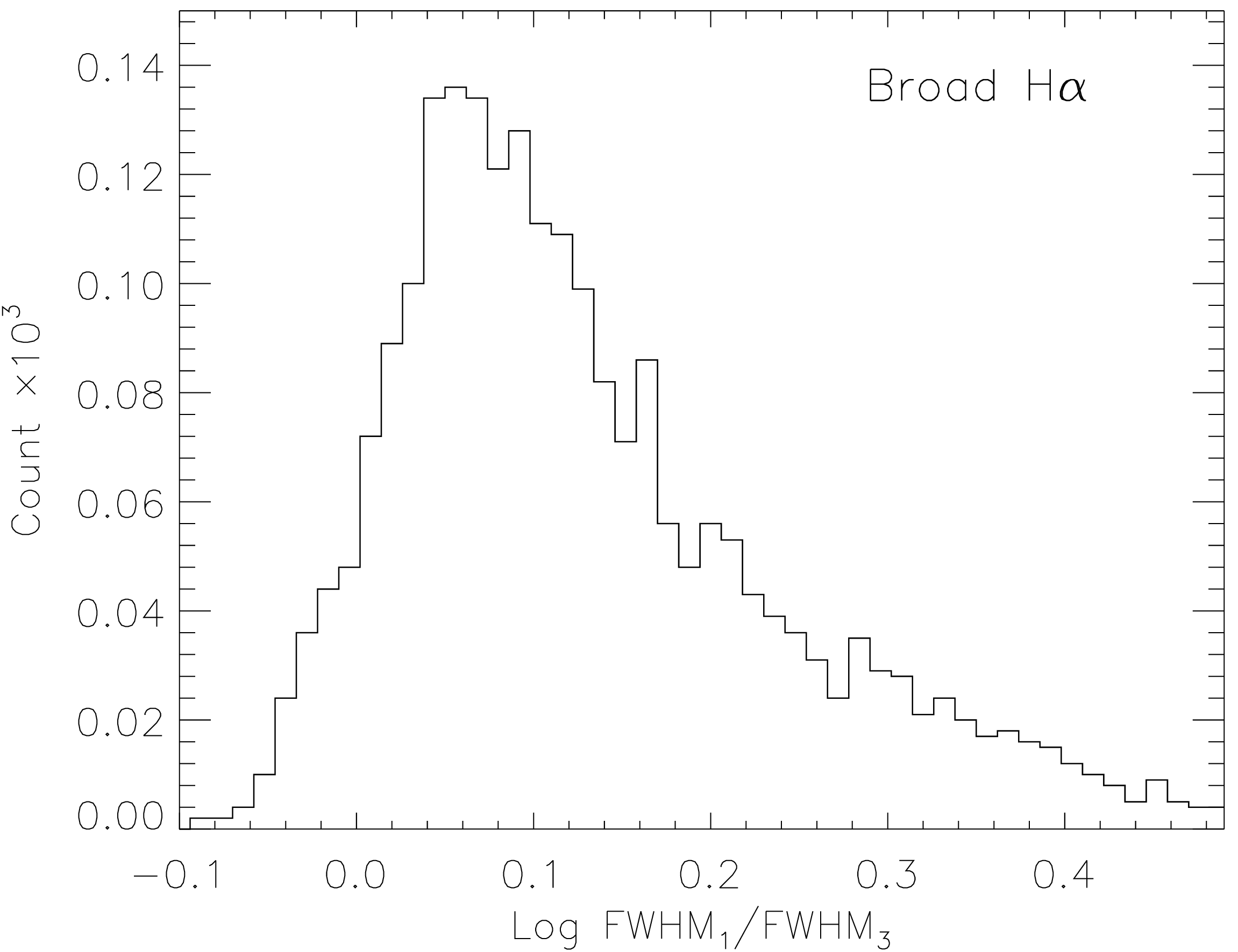}
\hspace{1ex}
\includegraphics[width=0.32\textwidth]{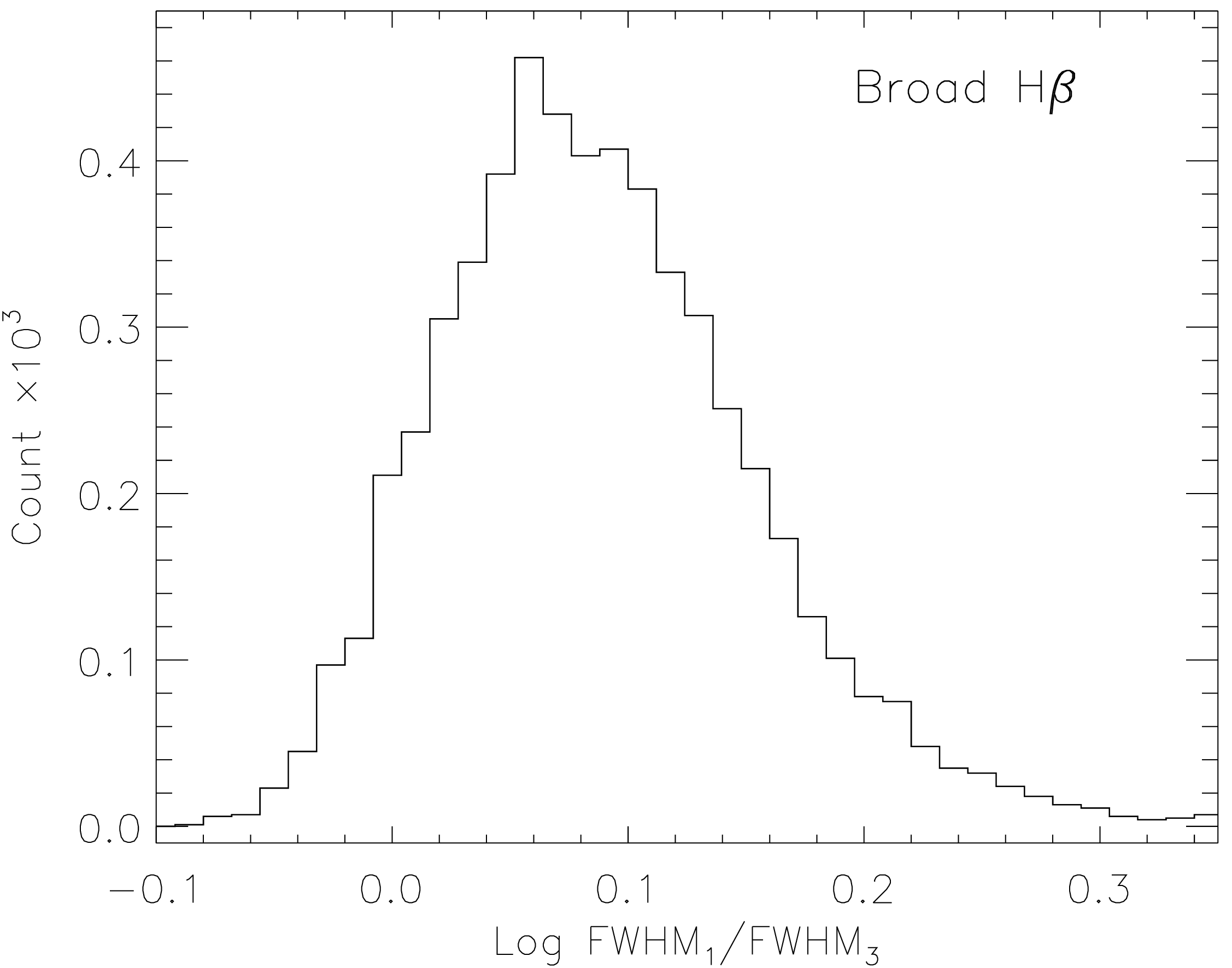}
\hspace{1ex}
\includegraphics[width=0.32\textwidth]{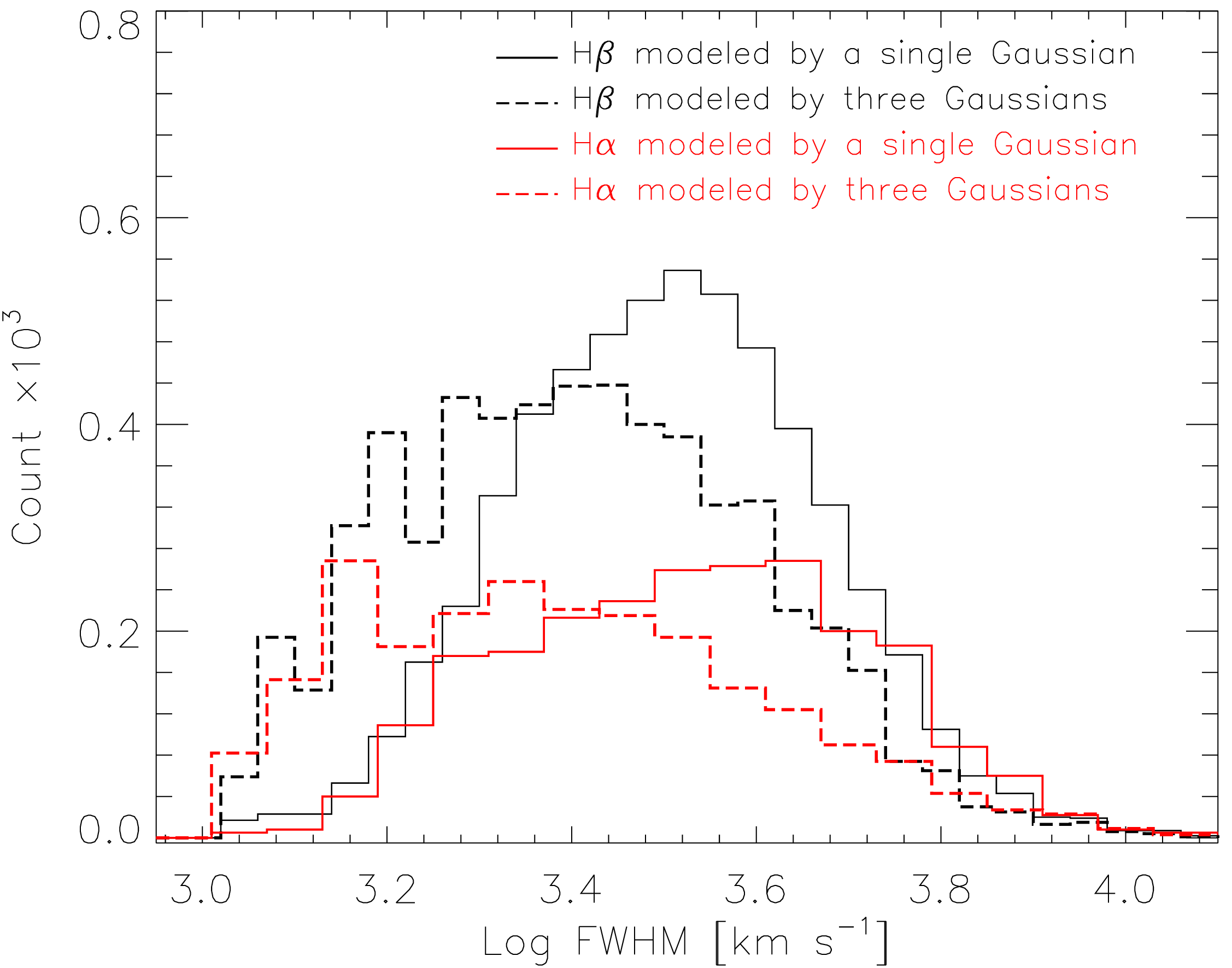}
\caption{Distributions of the FWHMs. The $\rm FWHM_1$ means the broad component modeled by a single Gaussian function, and the $\rm FWHM_3$ means the broad component modeled by three Gaussian functions. Left panel is for broad \halpha, and middle one is for broad \hbeta. Right panel: black lines are for broad \hbeta, red lines are for broad \halpha, solid-lines represent the broad component modeled by a single Gaussian function, and dash-lines represent the broad component modeled by three Gaussian functions.}
\label{fig:fwhm13}
\end{figure*}

\begin{figure*}
\centering
\includegraphics[width=0.32\textwidth]{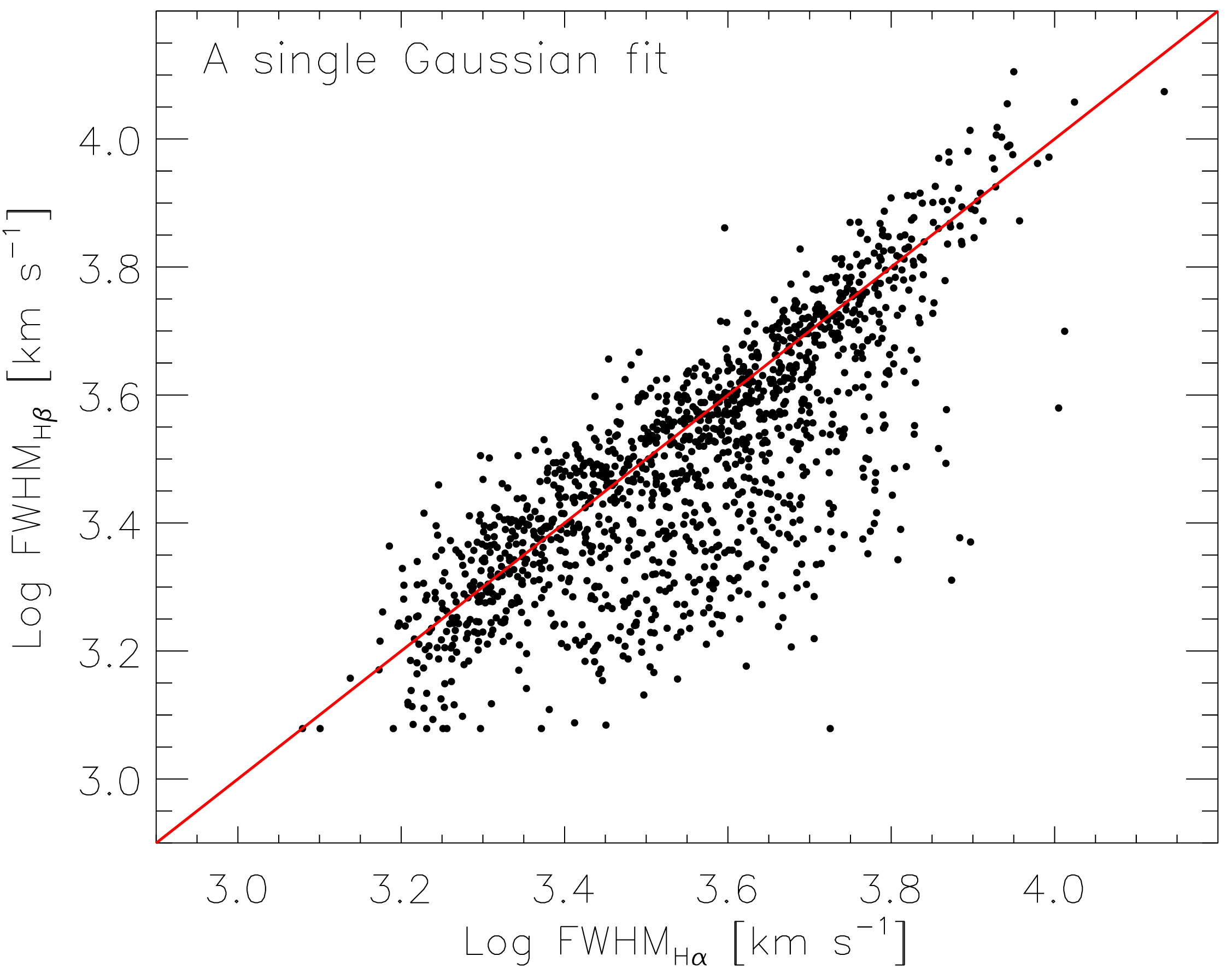}
\hspace{1ex}
\includegraphics[width=0.32\textwidth]{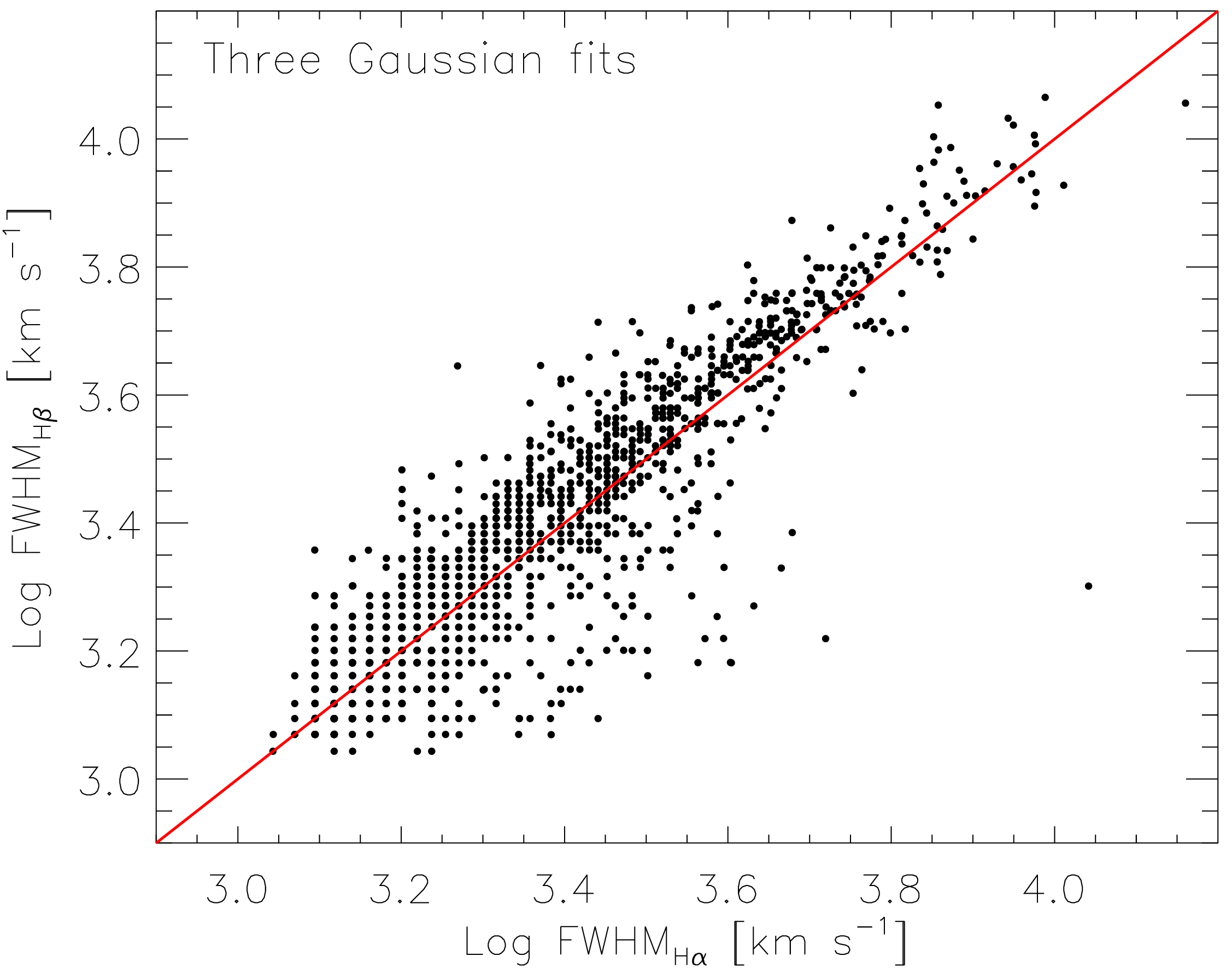}
\hspace{1ex}
\includegraphics[width=0.32\textwidth]{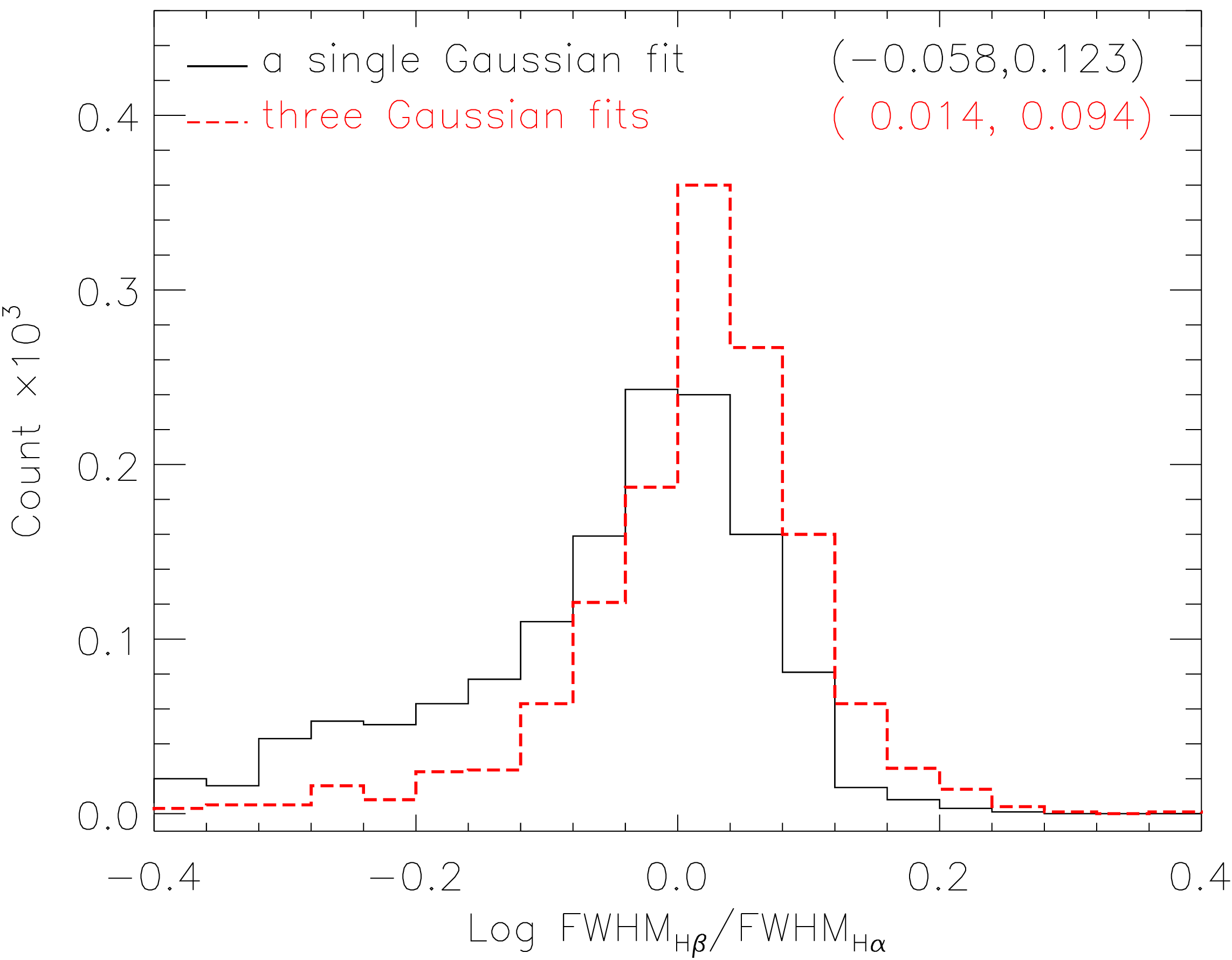}
\caption{Comparisons of FWHMs between \hbeta\ and \halpha. Red solid-lines indicate the identical values. Left panel: the broad \hbeta\ and \halpha\ components are modeled with a single Gaussian function. Middle panel: the broad \hbeta\ and \halpha\ components are modeled with three Gaussian functions. Right panel: ratios of FWHMs between \hbeta\ and \halpha. The values in the top-right corner indicate the mean offset and the standard deviation.}
\label{fig:fwhmab}
\end{figure*}

We model the broad \hbeta\ and \halpha\ components with either a single Gaussian function or a combination of three Gaussian functions. Figure \ref{fig:fwhm13} compares the FWHMs that are yielded by these two different methods. We find that a single Gaussian function fittings often yield a larger FWHMs relative to the three Gaussian function fitting. Nevertheless, we note that the differences between the FWHMs yielded by the two different methods would not produce a large bias in the virial mass of the black holes (less than 2.5 times).

Figure \ref{fig:fwhmab} compares the FWHMs between the broad \hbeta\ and \halpha\ for the quasars with both lines being available in the BOSS spectra. We find a strong correlation between the FWHMs of the broad \hbeta\ and \halpha, which is similar to previous studies \cite[e.g.;][]{2005ApJ...630..122G,2011ApJS..194...45S}. Nonetheless, it is clear that there is a large scatter if the broad \hbeta\ and \halpha\ are modeled with a single Gaussian function.

\begin{figure*}
\centering
\includegraphics[width=0.45\textwidth]{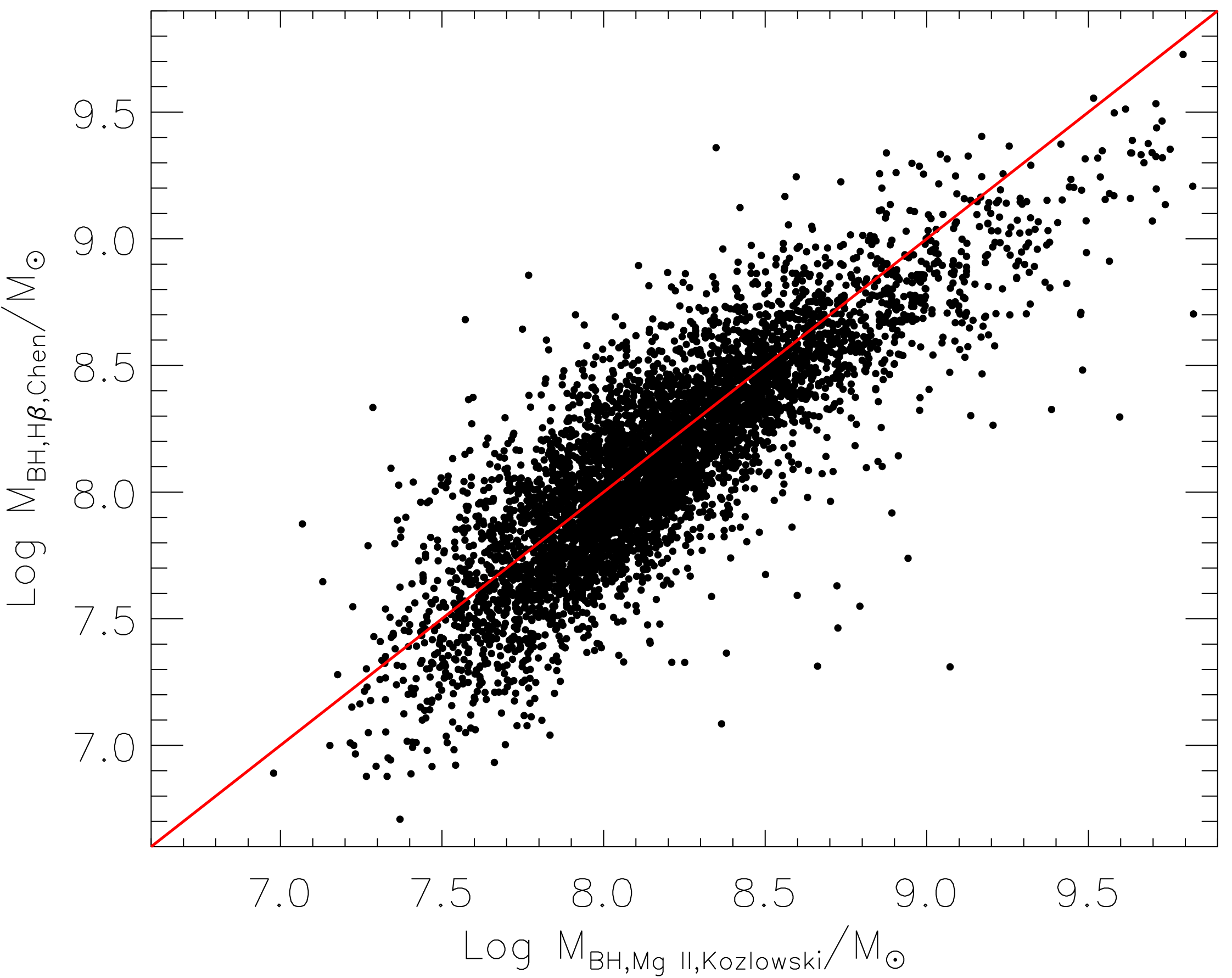}
\hspace{5ex}
\includegraphics[width=0.45\textwidth]{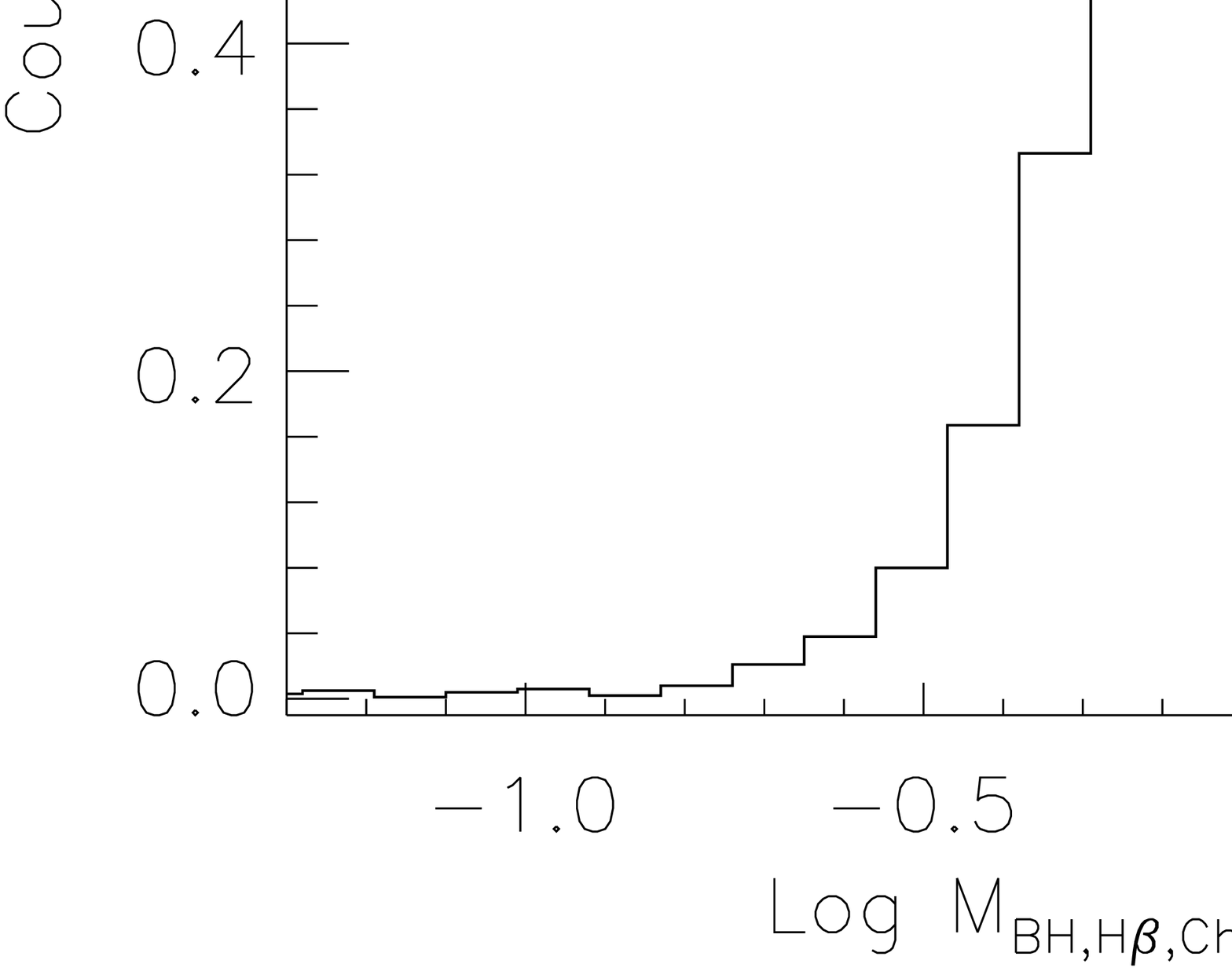}
\caption{Comparisons of the virial masses of the black holes. The FWHMs used to derive the virial mass are come from the fits modeling the broad emission components by a single Gaussian function. Left panel: the y-axis indicates our $M_{\rm BH}$ that are measured from broad \hbeta, and the x-axis indicates the $M_{\rm BH}$ of \cite{2017ApJS..228....9K} which are inferred from broad \MgII. The red solid line represents the identical values. Right panel: the $M_{\rm BH}$ ratio. The values in the top right corner indicate the mean offset and the standard deviation.}
\label{fig:MBHck}
\end{figure*}

For the quasars included in the DR12Q, \cite{2017ApJS..228....9K} inferred the continuum luminosities at 3000 \AA\ or 1350 \AA\ from the broadband urgiz magnitudes of the SDSS, and estimated the virial mass of the black holes for the quasars with broad \MgII\ or \CIV\ being available in the BOSS spectra, where the broad \MgII\ or \CIV\ emission components were modeled with a single Gaussian function \cite[e.g.;][]{2012AJ....144..144B,2017A&A...597A..79P}. For comparisons, we firstly derive the \hbeta\-based $M_{\rm BH}$ from Equation (\ref{eq:BHmass_Hbeta}) with calibrated coefficients of $\rm (a,b)=(6.91,0.50)$ \cite[e.g.;][]{2006ApJ...641..689V,2011ApJS..194...45S}, where the broad emission components are modeled with a single Gaussian function. In Figure \ref{fig:MBHck}, we compare the $M_{\rm BH}$ between the broad \MgII\ and \hbeta\ mass estimators for the quasars whose broad \MgII\ and \hbeta\ are simultaneously available in the BOSS spectra. The \MgII\ and \hbeta\ mass estimators are correlated with each other and yield consistent $M_{\rm BH}$ over three orders of magnitude in mass. The $M_{\rm BH}$ ratios between the \MgII\-based values and the \hbeta\-based value yield a negligible mean value of 0.053 dex and standard deviation of 0.239.

\begin{figure*}
\centering
\includegraphics[width=0.32\textwidth]{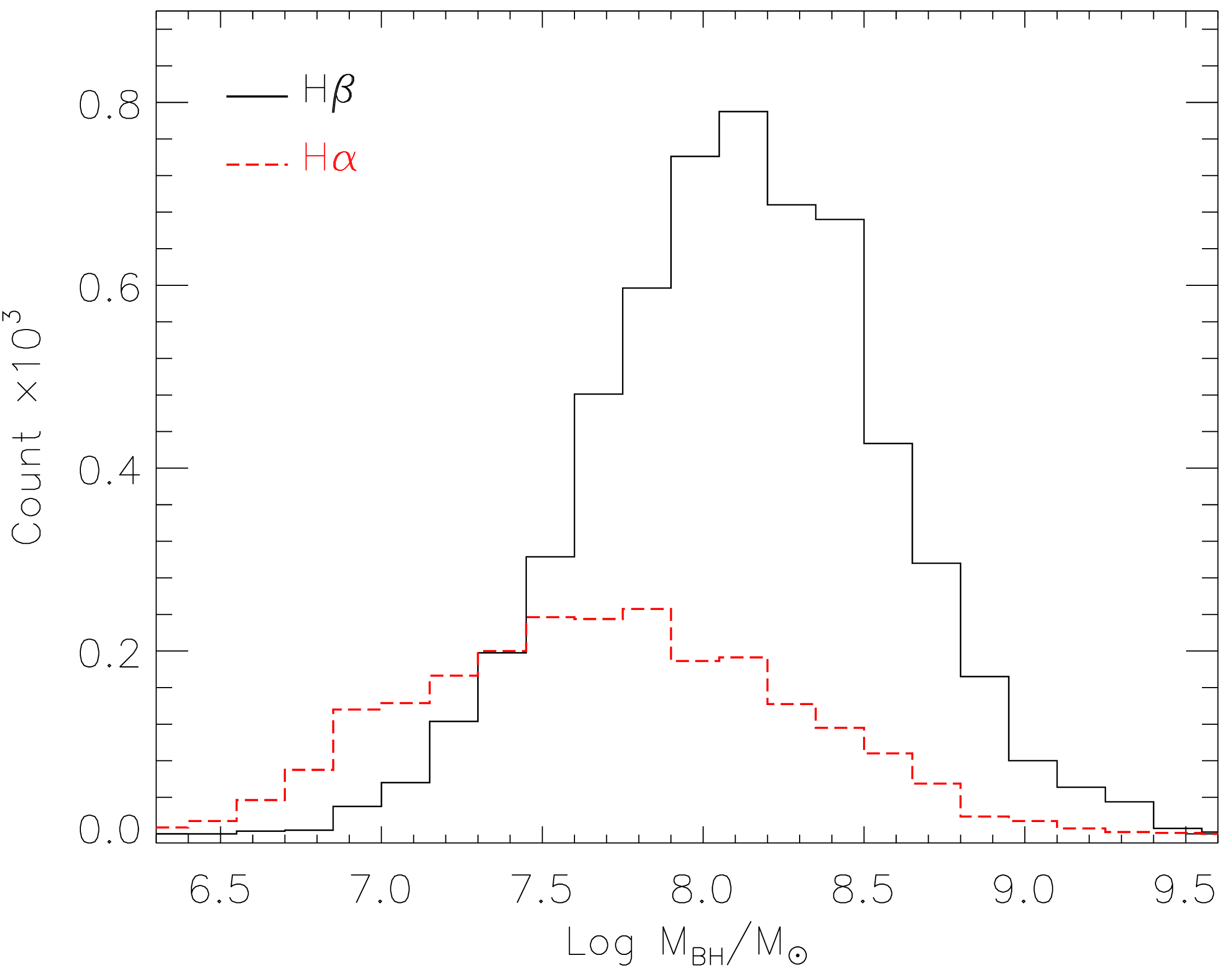}
\hspace{1ex}
\includegraphics[width=0.32\textwidth]{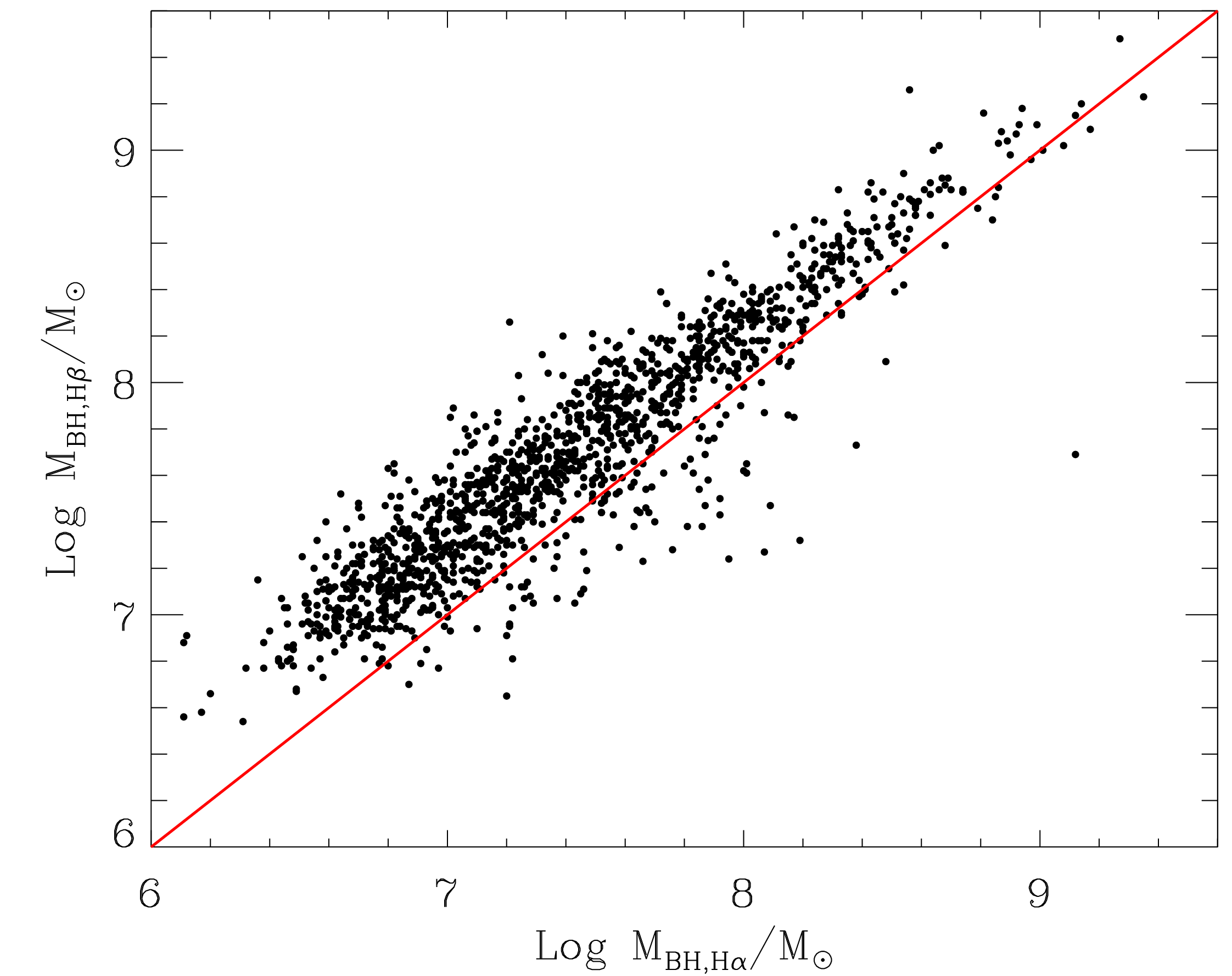}
\hspace{1ex}
\includegraphics[width=0.32\textwidth]{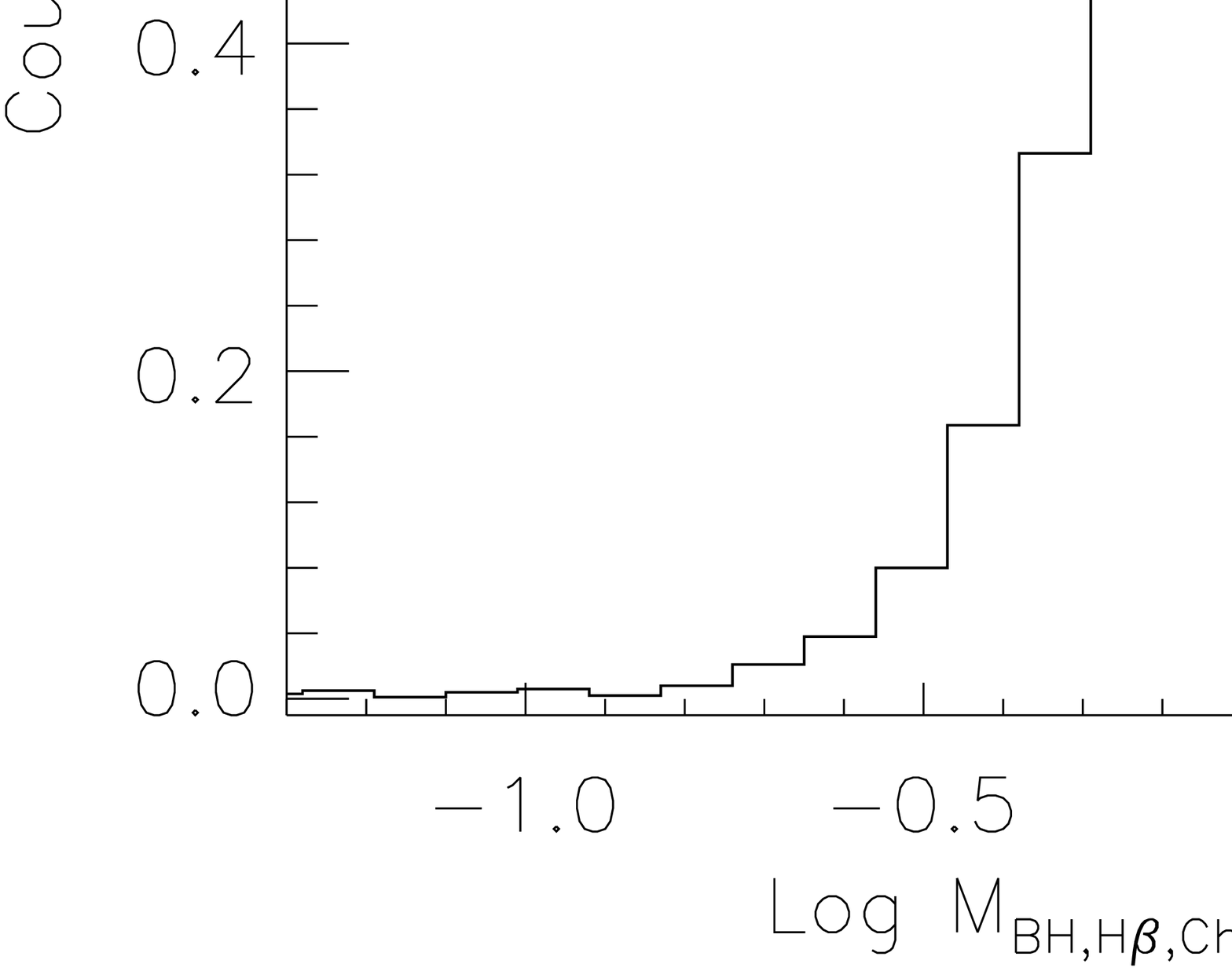}
\caption{The virial mass of the black holes are inferred from Equation (\ref{eq:BHmass_Hbeta}) or (\ref{eq:BHmass_Halpha}) with the calibrated coefficients in the literatures \cite[e.g.;][]{2005ApJ...630..122G,2006ApJ...641..689V,2011ApJS..194...45S}. Left panel: distributions of the $M_{\rm BH}$ for all quasars. The black solid-line represents the $M_{\rm BH}$ derived from broad \hbeta\ if they are available in the BOSS spectra, and red dash-line represents the $M_{\rm BH}$ derived from broad \halpha\ if they are available in the BOSS spectra. Middle panel: the comparison of the $M_{\rm BH}$ between \hbeta\ and \halpha\ line estimators for the quasars with both \hbeta\ and \halpha\ being available in the BOSS spectra. Red solid-line indicates the identical values. Right panel: the ratios of $M_{\rm BH}$. The values in the top-left corner indicate the mean offset and the standard deviation.}
\label{fig:MBHab}
\end{figure*}

In the above discussions, we noted that a combination of three Gaussian functions can better describe the broad emission components relative to a single Gaussian function. Therefore, in the following discussions, we calculate the virial mass of the black holes using the FWHMs of the fits modeling the broad \hbeta\ or \halpha\ components by a combination of three Gaussian functions. Both the continuum/line luminosities and the FWHMs are provided in our catalogs. Thus, readers can also infer the $M_{\rm BH}$ using the FWHMs of the fits modeling the broad \hbeta\ or \halpha\ components by a single Gaussian function. In the left panel of Figure \ref{fig:MBHab}, we show the distributions of the $M_{\rm BH}$ for all quasars, which are inferred from the broad \hbeta\ or \halpha\ if they are available in the BOSS spectra. It is clear that the $M_{\rm BH}$ of quasars in our sample are over three orders of magnitude. The middle and right panels of Figure \ref{fig:MBHab} compare the $M_{\rm BH}$ for the quasars whose broad \hbeta\ and \halpha\ are simultaneously available in the BOSS spectra. We find that there is a clear systematic difference between the $M_{\rm BH}$ inferred by the \hbeta\ and \halpha\ mass estimators. The ratios of the $M_{\rm BH}$ between the \hbeta\ and \halpha\ mass estimators show a significant offset of $0.269$ dex and a standard deviation of $0.22$.

\begin{figure*}
\centering
\includegraphics[width=0.45\textwidth]{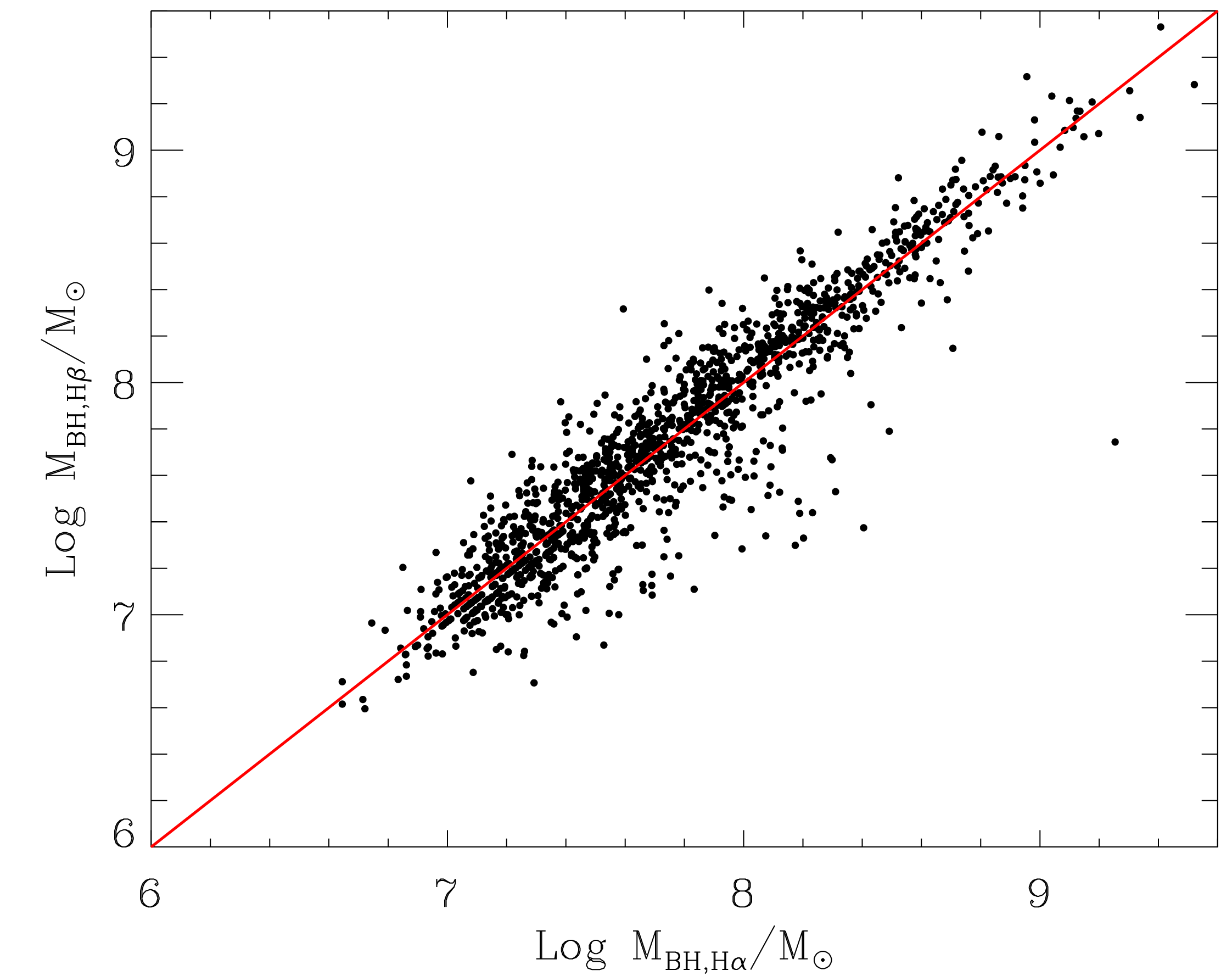}
\hspace{4ex}
\includegraphics[width=0.45\textwidth]{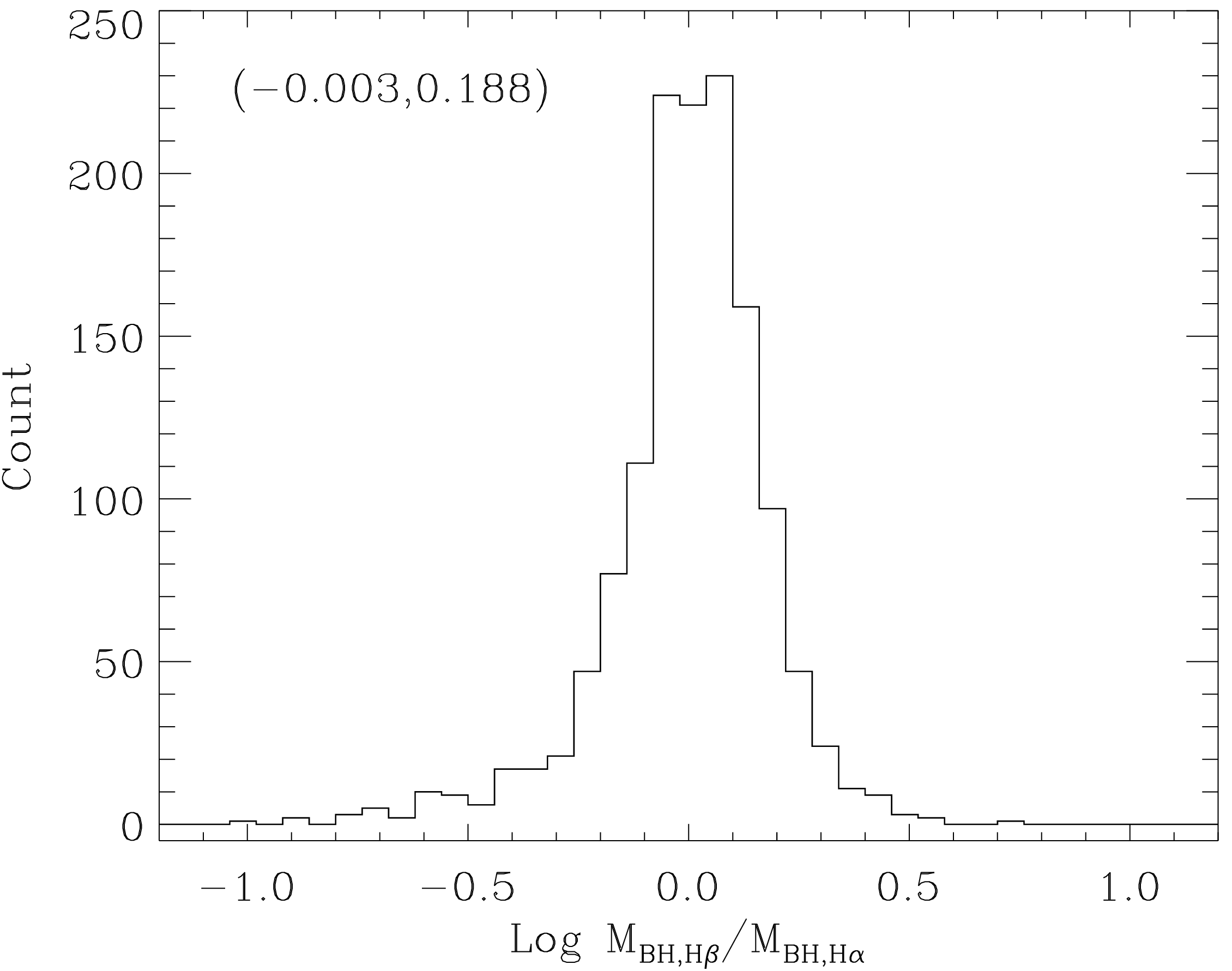}
\caption{The comparisons of the \hbeta\- and \halpha\-based mass, which are inferred from the Equation (\ref{eq:BHmass_Hbeta}) with the improved parameters of $\rm (a,b)=(7.00,0.50)$ for the \halpha\ and $\rm (a,b)=(6.96,0.50)$ for the \hbeta. The ed solid line indicates the identical values. The values in the top left corner indicate the mean offset and the standard deviation.}
\label{fig:MBHabnew}
\end{figure*}

We note that the \hbeta\- and \halpha\-based mass estimators are calibrated from much smaller samples \cite[e.g.;][]{2005ApJ...630..122G,2006ApJ...641..689V} compared to the quasar sample of this work. This could be an important reason why there is an obvious difference between the \hbeta\- and \halpha\-based masses. Here, using our large quasar sample, we improve the \hbeta\ and \halpha\ based mass estimators by minimizing the difference between the \hbeta\ and \halpha\ based mass. This requires two sets of parameters (a,b) to fix the Equation (\ref{eq:BHmass_Hbeta}), one for \hbeta\ mass estimator and one for \halpha\ mass estimator. We First fix the "b" parameter to be equal to 0.5, which is the theoretical expectation \cite[e.g.;][]{1997iagn.book.....P}, then select the "a" parameter on the (a,0.50) line. We find that $\rm (a,b)=(7.00,0.50)$ for the \halpha\ and $\rm (a,b)=(6.96,0.50)$ for the \hbeta\ are the best choices. Using these two sets of parameters, we infer the $M_{\rm BH}$ from the Equation (\ref{eq:BHmass_Hbeta}), respectively. The results are shown in Figure \ref{fig:MBHabnew}. The ratio between the \hbeta\ and \halpha\ based mass exhibits a negligible offset of $-0.003$ dex and a standard deviation of $0.188$.

\begin{figure*}
\centering
\includegraphics[width=0.32\textwidth]{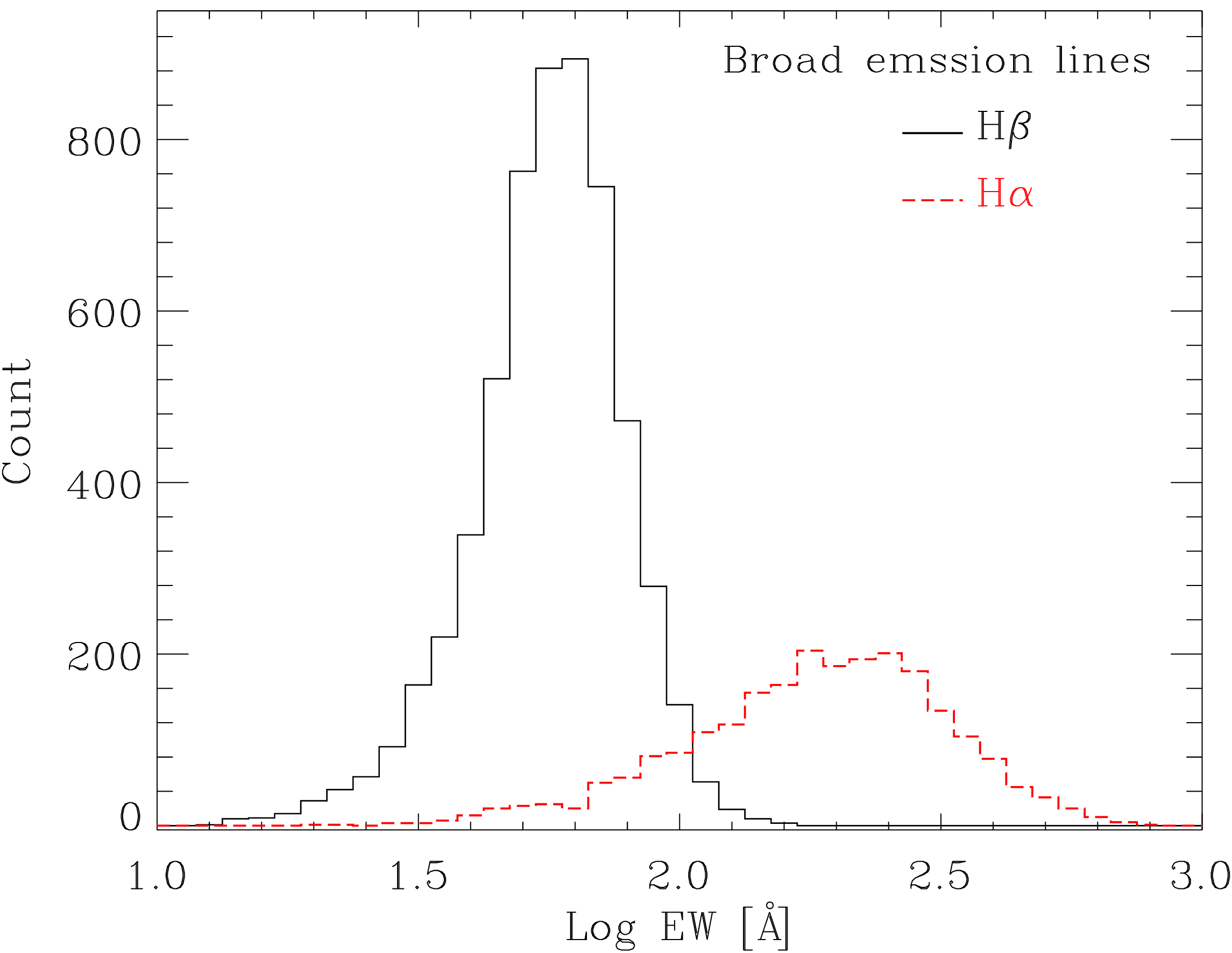}
\hspace{1ex}
\includegraphics[width=0.32\textwidth]{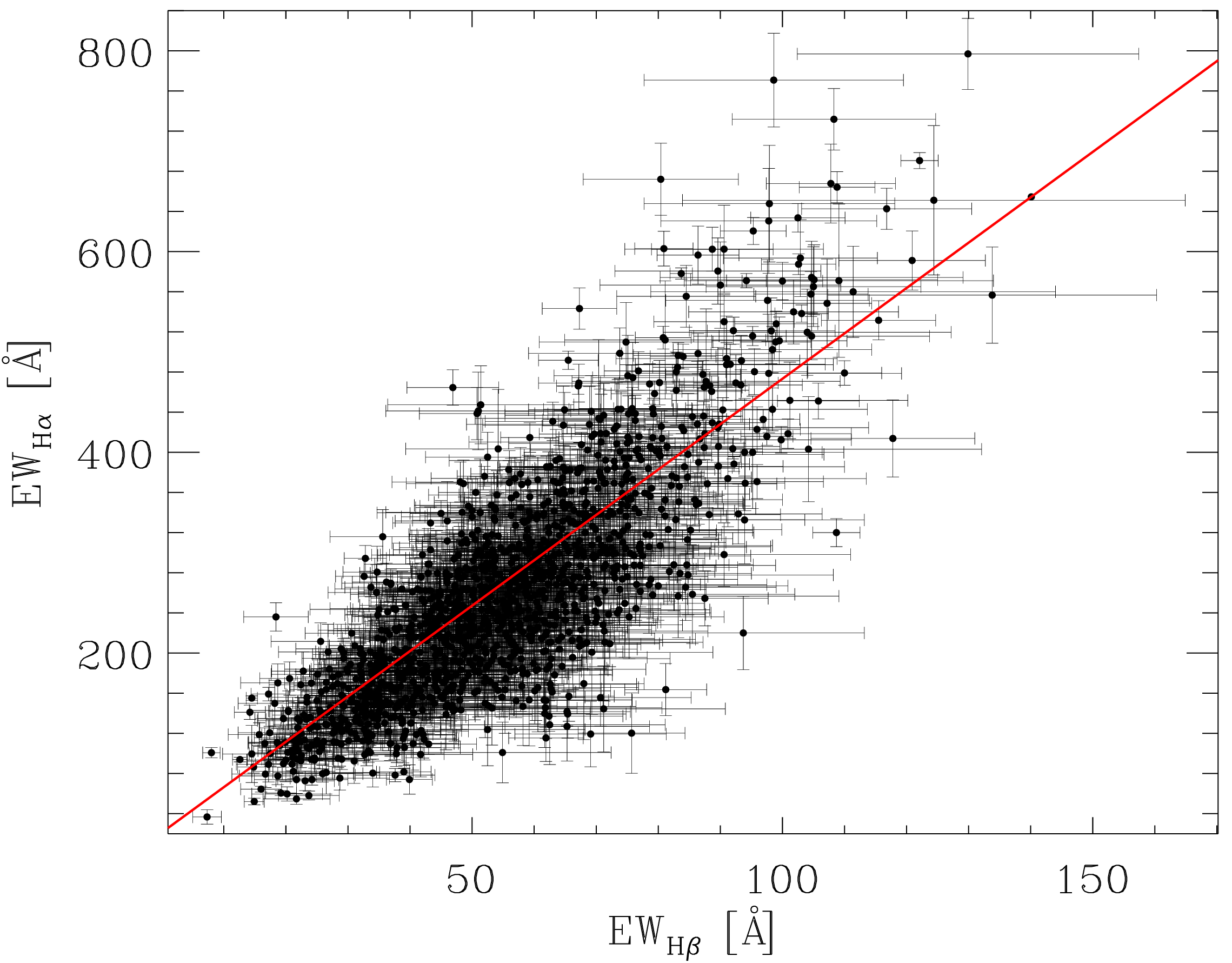}
\hspace{1ex}
\includegraphics[width=0.32\textwidth]{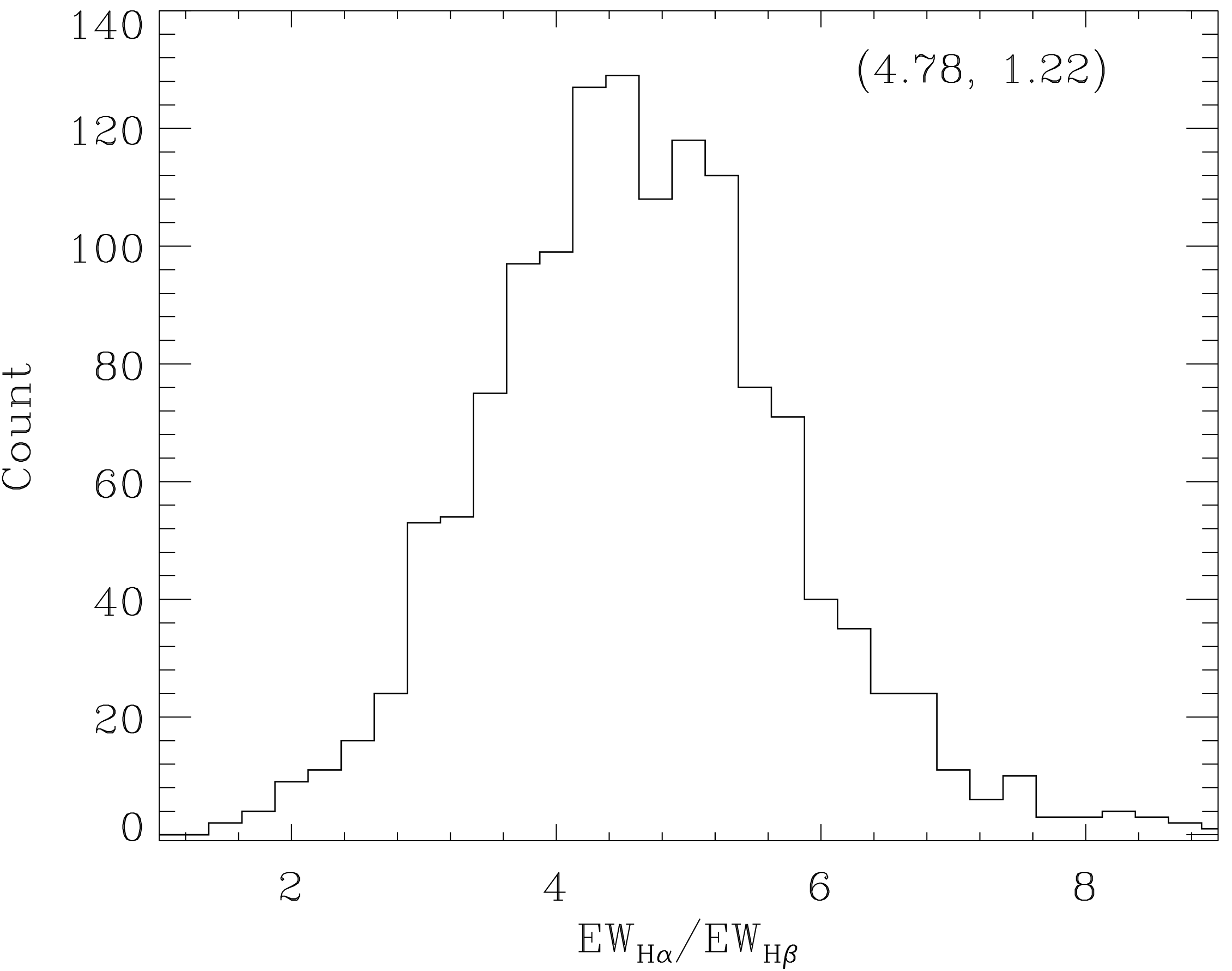}
\caption{Distributions of equivalent widths at rest-frame for emission lines with $EW_{\rm \lambda} \ge3\sigma_{EW_{\rm \lambda}}$. The left panel is for all quasars whose broad \hbeta\ (black solid line) or \halpha\ (red dash line) emission lines are available in the BOSS spectra. The middle and right panels are the comparisons of $EW_{\rm \lambda}$ between broad \hbeta\ and \halpha\ for the quasars whose broad \hbeta\ and \halpha\ are simultaneously available in the BOSS spectra. The red solid line is the ordinary least-square linear-fitting. The values in the top right corner indicate the mean offset and the standard deviation.}
\label{fig:ewab}
\end{figure*}

Figure \ref{fig:ewab} exhibits distributions of equivalent widths at rest-frame for emission lines with $EW_{\rm \lambda} \ge3\sigma_{EW_{\rm \lambda}}$. Both broad \hbeta\ and \halpha\ have equivalent widths over more than one order of magnitude. There is a remarkable correlation in the equivalent widths between broad \hbeta\ and \halpha. The ordinary least-square linear-fitting yields
\begin{equation}\label{eq:Lhbeta_Lhalpta}
\rm EW_{H\alpha} = (21.198 \pm 0.382) + (4.520 \pm 0.007) \times EW_{H\beta} ,
\end{equation}
which is shown as the red solid line in the middle panel of Figure \ref{fig:ewab}. In addition, broad \halpha\ are significantly stronger than broad \hbeta. The ratios of equivalent widths between broad \halpha\ and \hbeta\ show a mean value of 4.78 and a standard deviation of 1.22.

\begin{figure}
\centering
\vspace{6ex}
\includegraphics[width=0.45\textwidth]{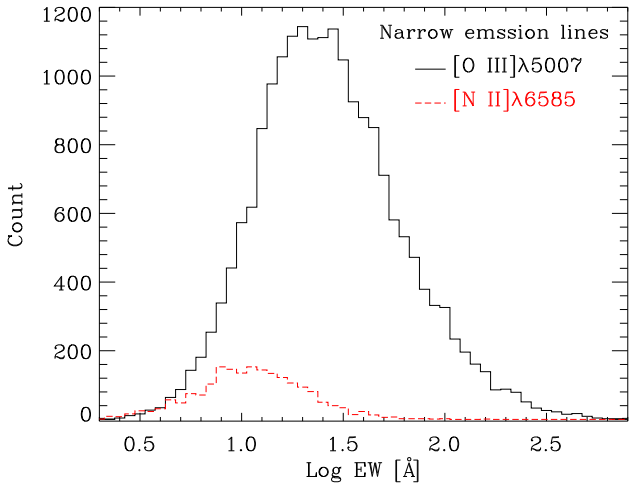}
\caption{Distributions of equivalent widths at rest-frame for the narrow \OIIIb\ (black solid line) and \NIIb\ (red dash line) emission lines with $EW_{\rm \lambda} \ge2\sigma_{EW_{\rm \lambda}}$.}
\label{fig:ewON}
\end{figure}

Narrow \NIIb\ emission lines have equivalent widths over more than one order of magnitude, while the narrow \OIIIb\ emission lines show a wider equivalent width range that is over two orders of magnitude. The results are exhibited in Figure \ref{fig:ewON}.

\begin{figure*}
\centering
\includegraphics[width=0.32\textwidth]{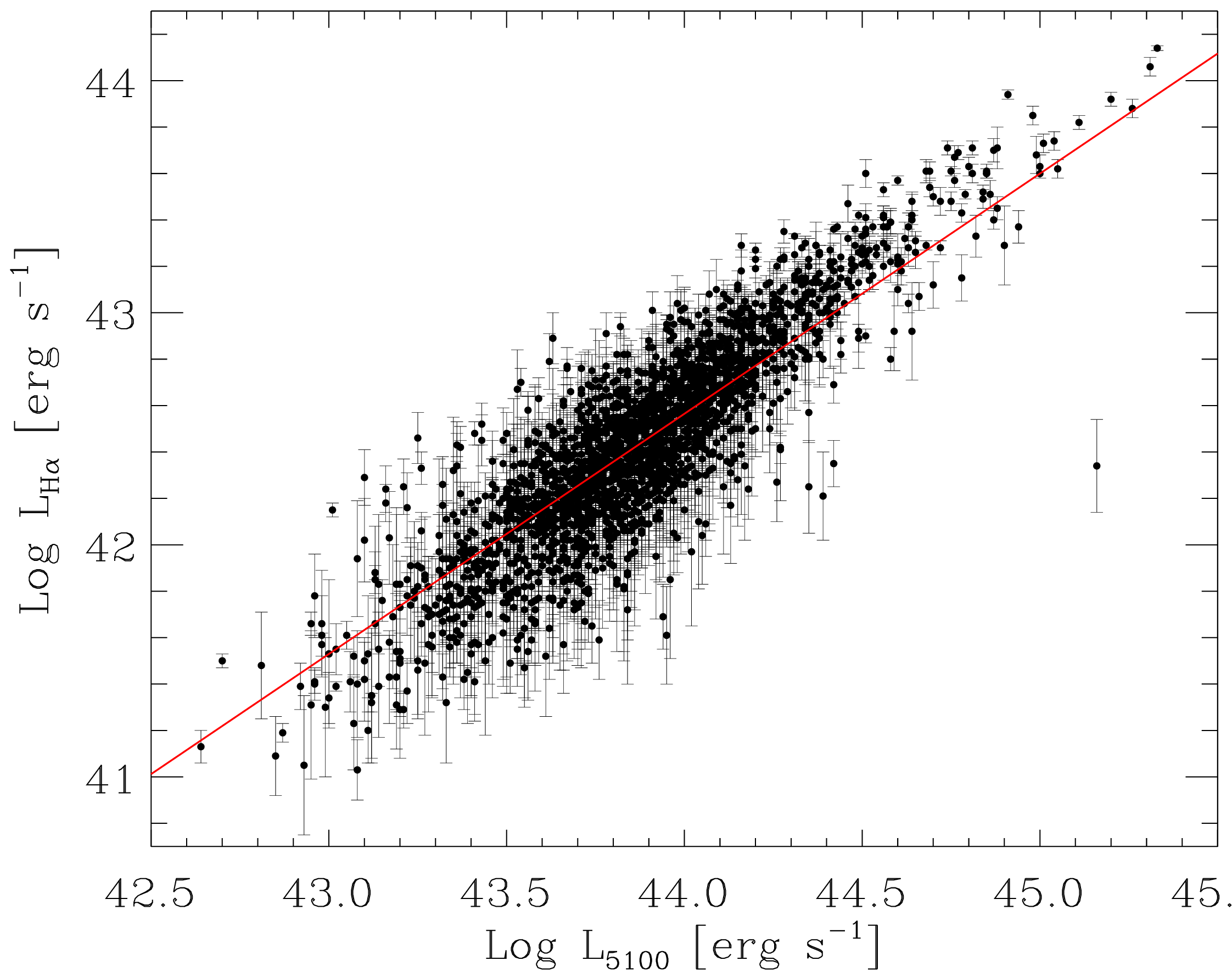}
\hspace{1ex}
\includegraphics[width=0.32\textwidth]{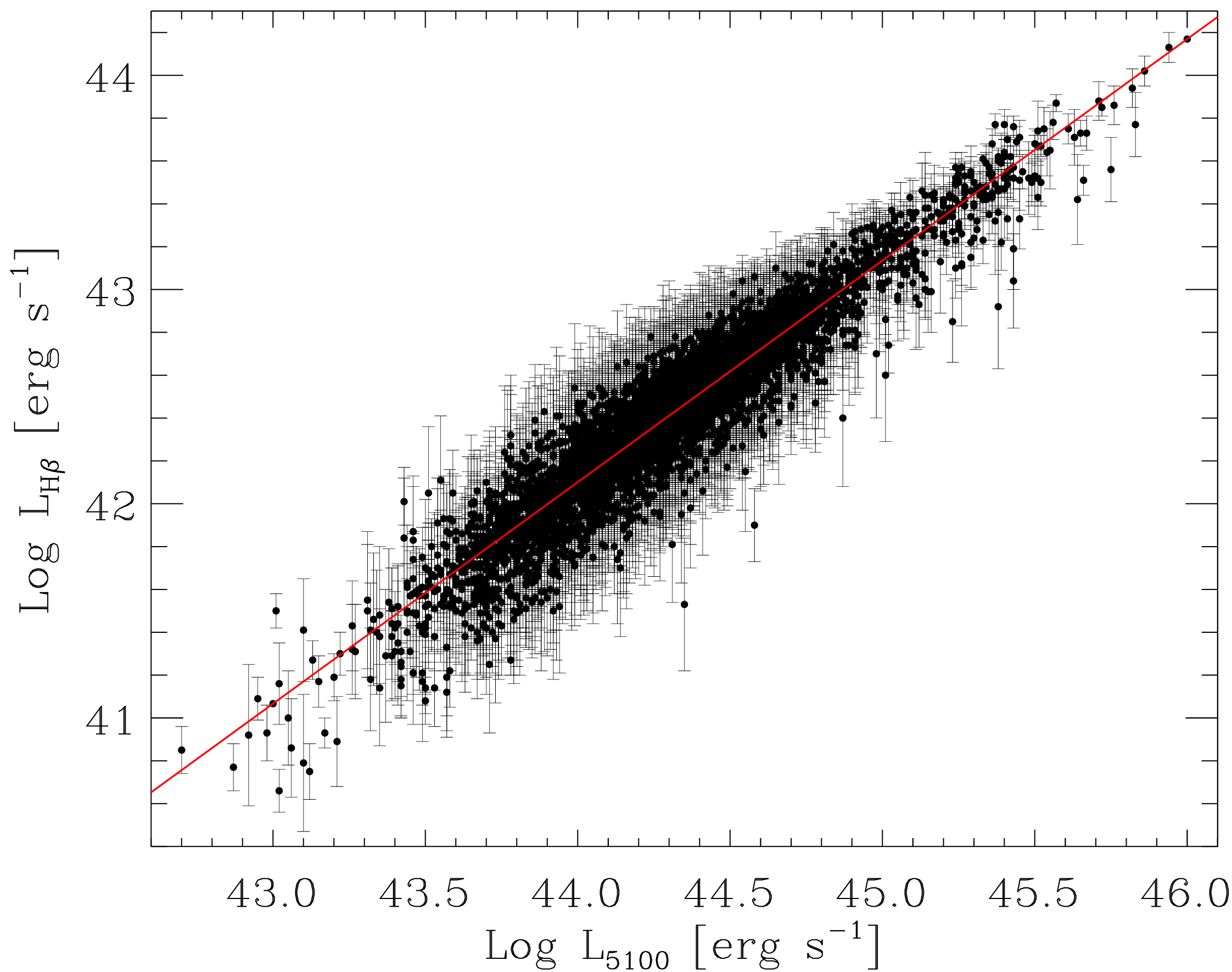}
\hspace{1ex}
\includegraphics[width=0.32\textwidth]{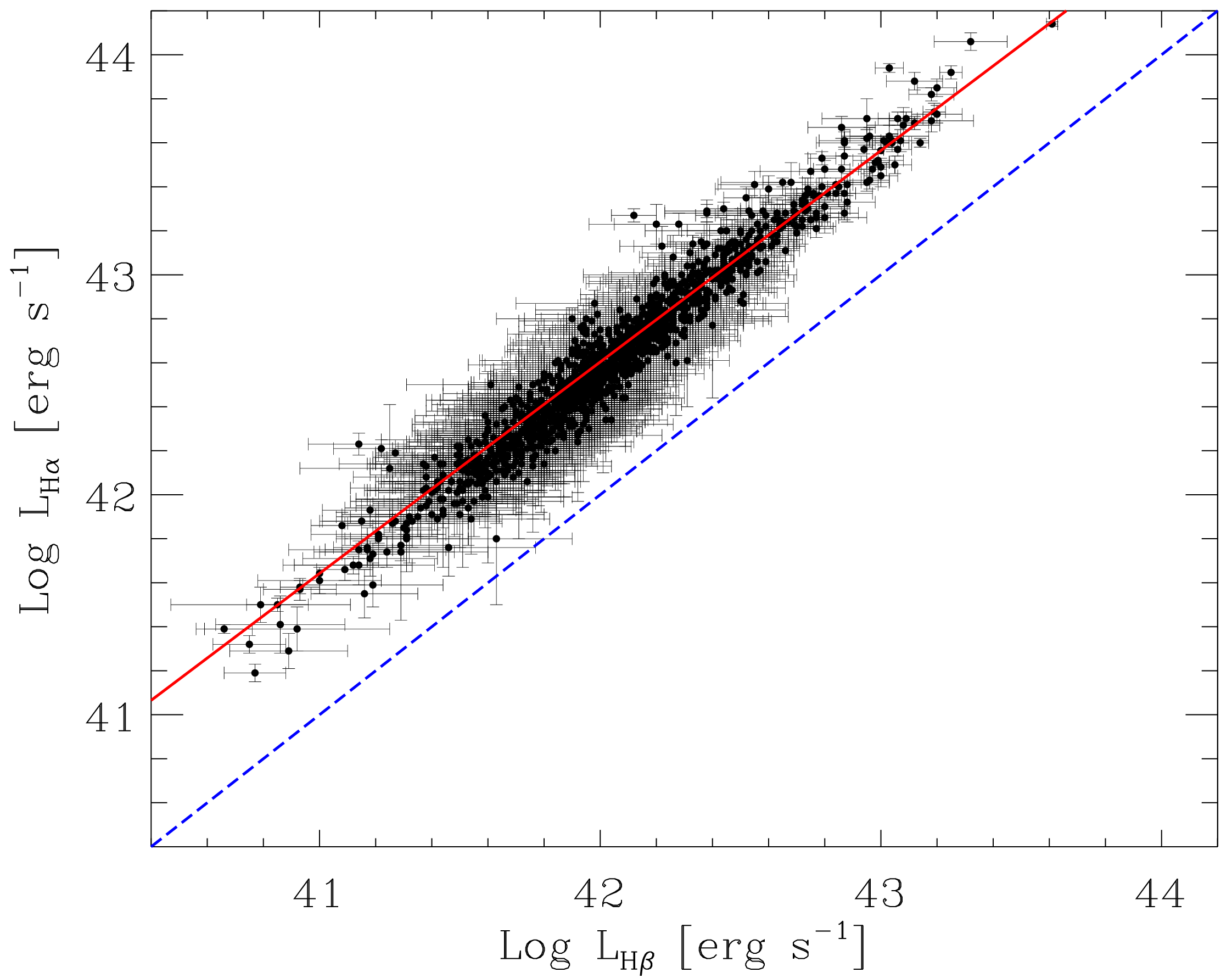}
\caption{Relationships of luminosities. The line luminosities are the total emissions of the broad and narrow components. The red solid lines indicate the ordinary least-squares linear fits. Left panel: \halpha\ line luminosity ($L_{\rm H\alpha}$) against continuum luminosity at 5100 \AA\ ($L_{\rm 5100}$) for the quasars whose broad \halpha\ are available in the BOSS spectra. Middle panel: \hbeta\ line luminosity ($L_{\rm H\beta}$) against $L_{\rm 5100}$ for the quasars whose broad \hbeta\ are available in the BOSS spectra. Right panel: $L_{\rm H\beta}$ against $L_{\rm H\alpha}$ for the quasars whose broad \hbeta\ and \halpha\ are simultaneously available in the BOSS spectra. The blue dash line indicates an identical value.}
\label{fig:LL}
\end{figure*}

The tight relationship between optical continuum luminosity and the luminosity of Balmer emission lines was uncovered about four decades ago \cite[e.g.;][]{1980ApJ...241..894Y,1981ApJ...244...12S}, and now the continuum luminosity is widely utilized as the proxy of \halpha\ and \hbeta\ line luminosities to characterize the properties of central objects. In Figure \ref{fig:LL}, we exhibit the relationships between continuum luminosity at 5100 \AA\ and the luminosity of Balmer emission lines for the quasars whose broad \hbeta\ and/or \halpha\ are available in the BOSS spectra. Here, the line luminosities are the total emissions of both the broad and narrow components. It is clear that both the \hbeta\ and \halpha\ line luminosities are significantly related with the continuum luminosity over three orders of magnitude in luminosity. The ordinary least-squares linear fits yield
\begin{equation}\label{eq:Lhalpta_L5100}
\rm Log L_{H\alpha} = (-2.971 \pm 0.606) + (1.035 \pm 0.014) \times Log L_{5100}
\end{equation}
for the \halpha, and
\begin{equation}\label{eq:Lhbeta_L5100}
\rm Log L_{H\beta} = (-3.402 \pm 0.218) + (1.034 \pm 0.005) \times Log L_{5100}
\end{equation}
for the \hbeta. The fitting results are overplotted as a red solid lines in Figure \ref{fig:LL}. Both the \hbeta\ and \halpha\ emissions trace the continuum luminosity with slopes that are consistent with unity and that are also similar to previous studies \cite[e.g.;][]{1981ApJ...244...12S,2005ApJ...630..122G}. In addition, there also is a tight correlation between the \halpha\ and \hbeta\ luminosities. The ordinary least-square linear fit yields
\begin{equation}\label{eq:Lhbeta_Lhalpta}
\rm Log L_{H\alpha} = (2.231 \pm 0.119) + (0.961 \pm 0.003) \times Log L_{H\beta} ,
\end{equation}
which is overplotted as a red solid-line in the right panel of Figure \ref{fig:LL}.

\section{summary}
\label{sect:summary}
In this paper, we mainly measure the spectral characteristics in the \hbeta\ and \halpha\ spectral regions for the low-redshift quasars included in DR12Q. The measurements in the \hbeta\ spectral region include emission lines of broad and narrow \hbeta, and \OIIIab\ doublets. The measurements in the \halpha\ spectral region include emission lines of broad and narrow \halpha, \NIIab\ and \SIIab\ doublets. We also derive the quasar continuum luminosity at 5100 \AA\ from the fitting power law. In addition, making use of empirical relationships, we infer the virial mass of black holes from broad \halpha\ and/or \hbeta. We also estimate quasar redshifts from \OIIIb\ emission lines if they are available in the BOSS spectra. These measured and derived quantities are available from the journal website.

The velocity offset of the visual inspection redshifts provided by \cite[][]{2017A&A...597A..79P} with respect to the \OIIIb\ based redshifts derived in this paper shows a negligible mean offset of 9 \kms\ and a dispersion of 92 \kms, which suggests that the visual inspection redshifts included in DR12Q should be robust for quasars with \zem\ $<0.9$.

For the quasars whose broad \hbeta\ and \halpha\ are available in the BOSS spectra, we find that the FWHMs, equivalent widths and line luminosities of the broad \hbeta\ are tightly correlated with the corresponding quantities of the broad \halpha. In addition, the FWHMs of the broad \halpha\ are consistent with those of the broad \hbeta, while both the equivalent widths and line luminosities of the broad \halpha\ are obviously larger than those of the broad \hbeta. These results are similar to previous studies.

We externally match the broad \hbeta\-based $M_{\rm BH}$ estimated in this paper to the broad \MgII\-based $M_{\rm BH}$ provided by \cite{2017ApJS..228....9K} for the quasars whose broad \hbeta\ and \MgII\ are available in the BOSS spectra, and find that the two mass estimators yield consistent $M_{\rm BH}$. We also cross-check internally the broad \hbeta\ and \halpha\ based $M_{\rm BH}$, and find that there is a significant offset between the \hbeta\- and \halpha\-based mass if they are inferred from the empirical relationships in the literature \cite[e.g.;][]{2005ApJ...630..122G,2006ApJ...641..689V}. Using our large quasar sample, we have improved the \hbeta\- and \halpha\-based mass estimators by minimizing the difference between the \hbeta\- and \halpha\-based mass. For the coefficients of (a,b) in the Equation (\ref{eq:BHmass_Hbeta}), we find that $\rm (a,b)=(7.00,0.50)$ for the \halpha\ and $\rm (a,b)=(6.96,0.50)$ for the \hbeta\ are the best choices. Using these two sets of parameters, the \hbeta\ and \halpha\ mass estimators produce consistent virial masses of the black holes.

\acknowledgements We much thank the anonymous referee for very helpful comments and structural suggestions. We acknowledge professors Qiu-Sheng Gu (Nanjing University, China) Yi-Ping Qin (Guangzhou University, China) for helpful conversations. This work was supported by the National Natural Science Foundation of China (NO. 11363001; NO. 11763001), the Guangxi Natural Science Foundation (2015GXNSFBA139004), the Excellent Youth Foundation of Guangdong Province (Grant No.YQ2015128), and the Guangzhou Education Bureau (Grant No.1201410593).

Funding for SDSS-III has been provided by the Alfred P. Sloan Foundation, the Participating Institutions, the National Science Foundation, and the U.S. Department of Energy Office of Science. The SDSS-III web site is http://www.sdss3.org/.

SDSS-III is managed by the Astrophysical Research Consortium for the Participating Institutions of the SDSS-III Collaboration including the University of Arizona, the Brazilian Participation Group, Brookhaven National Laboratory, Carnegie Mellon University, University of Florida, the French Participation Group, the German Participation Group, Harvard University, the Instituto de Astrofisica de Canarias, the Michigan State/Notre Dame/JINA Participation Group, Johns Hopkins University, Lawrence Berkeley National Laboratory, Max Planck Institute for Astrophysics, Max Planck Institute for Extraterrestrial Physics, New Mexico State University, New York University, Ohio State University, Pennsylvania State University, University of Portsmouth, Princeton University, the Spanish Participation Group, University of Tokyo, University of Utah, Vanderbilt University, University of Virginia, University of Washington, and Yale University.


\begin{table*}\tiny
\caption{Description of the catalog including the measurements around the \halpha\ spectral region.} \tabcolsep 1.7mm
\centering
\label{tab:halpha}
 \begin{tabular}{ll}
 \hline\hline\noalign{\smallskip}
 Column & Description\\
\hline\noalign{\smallskip}
      1 & SDSS DR12 designation (J2000.0; truncated coordinates): hhmmss.ss+ddmmss.s.\\
      2 & Spectroscopic plate number, Modified Julian Date (MJD), and spectroscopic fiber number: plateID-MJD-fiberID.\\
      3 & Redshift from the visual inspection, which is provided in DR12Q.\\
      4 & Redshift determined by $\rm [O~III]\lambda5007$ emission line. 0.0 indicates that the $\rm EW_{\OIII\lambda5007}<2\sigma_{\rm EW_{\OIII\lambda5007}}$.\\
      5 & Continuum luminosity at 5100 \AA, which is directly determined from the fitting power-law.\\
      6 & Black hole mass inferred by Equation (\ref{eq:BHmass_Halpha}) with the FWHM of the broad \halpha\ modeled by a combination of three Gaussian functions. 0.0 indicates that the broad \halpha\ has $\rm EW_{H\alpha}<3\sigma_{\rm EW_{H\alpha}}$.\\
      7 & Black hole mass inferred by Equation }{\ref{eq:BHmass_Hbeta}) with the $\rm (a,b)=(7.00,0.50)$ and the FWHM of the broad \hbeta\ modeled by a combination of three Gaussian functions. 0.0 indicates that the broad has $\rm EW_{H\beta}<3\sigma_{\rm EW_{H\beta}}$.\\
      8 & FWHM of the broad \halpha\ component modeled by a single Gaussian function. 0.0 indicates that the broad \halpha\ component has $\rm EW_{H\alpha}<3\sigma_{\rm EW_{H\alpha}}$.\\
      9 & FWHM of the broad \halpha\ component modeled by a combination of three Gaussian functions. 0.0 indicates that the broad \halpha\ component has $\rm EW_{H\alpha}<3\sigma_{\rm EW_{H\alpha}}$.\\
10 --- 13& Line luminosity, equivalent width, and corresponding $1\sigma$ errors for the broad \halpha\ modeled by a combination of three Gaussian functions. 0.0 indicates that the broad \halpha\ has $\rm EW_{ H\alpha}<3\sigma_{\rm EW_{H\alpha}}$.\\
      14 & Total line luminosity of the broad plus narrow \halpha\ components. 0.0 indicates that the broad \halpha\ component has $\rm EW_{H\alpha}<3\sigma_{\rm EW_{H\alpha}}$.\\
15 --- 18& Line luminosity, equivalent width, and corresponding $1\sigma$ errors for the narrow \halpha\ component. 0.0 indicates that the narrow \halpha\ component has $\rm EW_{H\alpha}<2\sigma_{\rm EW_{H\alpha}}$.\\
      19& FWHM of the $\rm [N~II]\lambda6585$. 0.0 indicates that the $\rm [N~II]\lambda6585$ has $\rm EW_{\NII\lambda6585}<2\sigma_{\rm EW_{\NII\lambda6585}}$.\\
20 --- 23& Line luminosity, equivalent width, and corresponding $1\sigma$ errors of the $\rm [N~II]\lambda6585$. 0.0 indicates that the $\rm EW_{\NII\lambda6585}<2\sigma_{\rm EW_{\NII\lambda6585}}$.\\
24 --- 27& Line luminosity, equivalent width, and corresponding $1\sigma$ errors of the $\rm [N~II]\lambda6549$. 0.0 indicates that the $\rm EW_{\NII\lambda6585}<2\sigma_{\rm EW_{\NII\lambda6585}}$.\\
28 --- 31& Line luminosity, equivalent width, and corresponding $1\sigma$ errors of the $\rm [S~II]\lambda6732$. 0.0 indicates that the $\rm EW_{\SII\lambda6732}<2\sigma_{\rm EW_{\SII\lambda6732}}$ or $\rm EW_{\SII\lambda6718}<2\sigma_{\rm EW_{\SII\lambda6718}}$.\\
32 --- 35& Line luminosity, equivalent width, and corresponding $1\sigma$ errors of the $\rm [N~II]\lambda6718$. 0.0 indicates that the $\rm EW_{\SII\lambda6732}<2\sigma_{\rm EW_{\SII\lambda6732}}$ or $\rm EW_{\SII\lambda6718}<2\sigma_{\rm EW_{\SII\lambda6718}}$.\\
       36& Reduced $\chi^2$ of the line fits in \halpha\ spectral region. The broad \halpha\ components are modeled with a combination of three Gaussian functions.\\
       37& Reduced $\chi^2$ of the line fits in \halpha\ spectral region. The broad \halpha\ components are modeled with a single Gaussian function.\\
38 --- 39& Number of good pixels and median S/N per pixel in \hbeta\ spectral region (6400 --- 6800 \AA).\\
\noalign{\smallskip}
\hline\hline\noalign{\smallskip}
\end{tabular}
\end{table*}

\begin{table*}\tiny
\caption{Description of the catalog including the measurements around the \hbeta\ spectral region.} \tabcolsep 1.7mm
\centering
\label{tab:hbeta}
 \begin{tabular}{ll}
 \hline\hline\noalign{\smallskip}
 Column & Description\\
\hline\noalign{\smallskip}
      1 & SDSS DR12 designation (J2000.0; truncated coordinates): hhmmss.ss+ddmmss.s.\\
      2 & Spectroscopic plate number, Modified Julian Date (MJD), and spectroscopic fiber number: plateID-MJD-fiberID.\\
      3 & Redshift from the visual inspection, which is provided in DR12Q.\\
      4 & Redshift determined by $\rm [O~III]\lambda5007$ emission line. 0.0 indicates that the $\rm EW_{\OIII\lambda5007}<2\sigma_{\rm EW_{\OIII\lambda5007}}$.\\
      5 & Continuum luminosity at 5100 \AA, which is directly determined from the fitting power-law.\\
      6 & Black hole mass inferred by Equation (\ref{eq:BHmass_Hbeta}) with the $\rm (a,b)=(6.91,0.50)$ and the FWHM of the broad \hbeta\ modeled by a combination of three Gaussian functions. 0.0 indicates that the broad has $\rm EW_{H\beta}<3\sigma_{\rm EW_{H\beta}}$.\\
      7 & Black hole mass inferred by Equation (\ref{eq:BHmass_Hbeta}) with the $\rm (a,b)=(6.96,0.50)$ and the FWHM of the broad \hbeta\ modeled by a combination of three Gaussian functions. 0.0 indicates that the broad has $\rm EW_{H\beta}<3\sigma_{\rm EW_{H\beta}}$.\\
      8 & FWHM of the broad \hbeta\ component modeledy by a single Gaussian function. 0.0 indicates that the broad \hbeta\ component has $\rm EW_{H\beta}<3\sigma_{\rm EW_{H\beta}}$.\\
      9 & FWHM of the broad \hbeta\ component modeled by a combination of three Gaussian functions. 0.0 indicates that the broad \hbeta\ component has $\rm EW_{H\beta}<3\sigma_{\rm EW_{H\beta}}$.\\
10 --- 13& Line luminosity, equivalent width, and corresponding $1\sigma$ errors for the broad \hbeta\ modeled by a combination of three Gaussian functions. 0.0 indicates that the broad \hbeta\ has $\rm EW_{ H\beta}<3\sigma_{\rm EW_{H\beta}}$.\\
      14 & Total line luminosity of the broad plus narrow \hbeta\ components. 0.0 indicates that the broad \hbeta\ component has $\rm EW_{H\beta}<3\sigma_{\rm EW_{H\beta}}$.\\
15 --- 18& Line luminosity, equivalent width, and corresponding $1\sigma$ errors for the narrow \hbeta\ component. 0.0 indicates that the narrow \hbeta\ component has $\rm EW_{H\beta}<2\sigma_{\rm EW_{H\beta}}$.\\
       19& FWHM of the narrow \hbeta\ component. 0.0 indicates that the narrow \hbeta\ component has $\rm EW_{H\beta}<2\sigma_{\rm EW_{H\beta}}$.\\
20 --- 23& Total line luminosity, equivalent width, and corresponding $1\sigma$ errors of the $\rm [O~III]\lambda5007$, which are the sum of the core and wing components. 0.0 indicates that the $\rm EW_{ \OIII\lambda5007}<2\sigma_{\rm EW_{\OIII\lambda5007}}$.\\
24 --- 27& Total line luminosity, equivalent width, and corresponding $1\sigma$ errors of the $\rm [O~III]\lambda4959$, which are the sum of the core and wing components. 0.0 indicates that the $\rm EW_{ \OIII\lambda5007}<2\sigma_{\rm EW_{\OIII\lambda5007}}$.\\
       28& Reduced $\chi^2$ of the line fits in \hbeta\ spectral region. The broad \hbeta\ components are modeled with a combination of three Gaussian functions.\\
       29&Reduced $\chi^2$ of the line fits in \hbeta\ spectral region. The broad \hbeta\ components are modeled with a single Gaussian function.\\
30 --- 31& Number of good pixels and median S/N per pixel in \hbeta\ spectral region (4700 --- 5100 \AA).\\
\noalign{\smallskip}
\hline\hline\noalign{\smallskip}
\end{tabular}
\end{table*}

\end{document}